\documentclass[12pt,preprint]{emulateapj}

\slugcomment{Draft Version May 24, 2012}

\shorttitle{Infrared Properties of Local AGNs}
\shortauthors{Ichikawa et al.}

\usepackage{lscape}
\usepackage{amsmath}
\usepackage{color}
\usepackage{url}
\usepackage{ulem}
\usepackage{multirow}

\begin{document}

\title{Mid and Far Infrared Properties of a Complete Sample of Local AGNs}


\author{Kohei Ichikawa\altaffilmark{1}, Yoshihiro Ueda\altaffilmark{1}, Yuichi Terashima\altaffilmark{2}, Shinki Oyabu\altaffilmark{3},
 Poshak Gandhi\altaffilmark{4}, Keiko Matsuta\altaffilmark{5}, and Takao Nakagawa\altaffilmark{4} }
\affil{\altaffilmark{1}
Department of Astronomy, Graduate School of Science, Kyoto University,
Kitashirakawa-Oiwake cho, Kyoto, 606-8502, Japan \\
\altaffilmark{2} 
Department of Physics, Faculty of Science, Ehime University, Matsuyama 790-8577, Japan\\
\altaffilmark{3}
Graduate School of Science, Nagoya University, Furo-cho, Chikusa-ku, Nagoya, Aichi 464-8602 Japan\\
\altaffilmark{4}
Institute of Space and Astronautical Science (ISAS), Japan Aerospace Exploration Agency, 3-1-1 Yoshinodai, Chuo-ku, Sagamihara, Kanagawa 252-5210, Japan\\
\altaffilmark{5}
Department of Space and Astronautical Science, The Graduate University for Advanced Studies, 3-1-1Yoshinodai, Chuo-ku, Sagamihara, Kanagawa 252-5210, Japan
}
\email{ichikawa@kusastro.kyoto-u.ac.jp}

\begin{abstract}

We investigate the mid- (MIR) to far-infrared (FIR) properties of a
nearly complete sample of local Active Galactic Nuclei (AGNs) detected
in the {\it Swift}/BAT all sky hard X-ray (14--195 keV) survey, based
on the cross correlation with the {\it AKARI} infrared survey catalogs
complemented by those with {\it IRAS} and {\it WISE}. Out of 135
non-blazer AGNs in the {\it Swift}/BAT 9 month catalog, we obtain the
MIR photometric data for 128 sources either in the 9, 12, 18, 22,
and/or 25 $\mu$m band. We find good correlation between their hard
X-ray and MIR luminosities over 3 orders of magnitude ($42< \log \lambda L_{\lambda}(9,
18~\mu{\rm m})< 45$), which is tighter than that with the FIR
luminosities at 90 $\mu$m. This suggests that thermal emission from
hot dusts irradiated by the AGN emission dominate the MIR fluxes. Both
X-ray unabsorbed and absorbed AGNs follow the same correlation,
implying isotropic infrared emission, as expected in 
clumpy dust tori rather than homogeneous ones.
We find excess signals around $9~\mu$m in the
averaged infrared spectral energy distribution from heavy obscured
``new type'' AGNs with small scattering fractions in the X-ray
spectra. This could be attributed to the PAH emission feature,
suggesting that their host galaxies have strong starburst activities.

\end{abstract}


\section{Introduction}

A complete survey of Active Galactic Nuclei (AGNs) throughout the
history of the universe is one of main goals in modern astronomy, which
is necessary to understand the evolution of supermassive black holes
(SMBHs) in galactic centers and their host galaxies. Given the fact that
the majority of AGNs are obscured by dust and gas surrounding the SMBH,
observations in hard X-ray and mid-infrared (MIR) bands are proposed to be
promising tools for detecting the whole populations of AGNs (both radio
quiet and loud ones) thanks to their strong penetrating power than
optical/UV lights and soft X-rays. In fact, recent deep
multi-wavelengths surveys utilizing these energy bands are discovering a
large number of obscured AGNs \citep{bra05}. Hard X-ray
selection gives the most efficient way to have a clean AGN sample with
little contamination from host galaxies. On the other hand, MIR
selection sometimes achieves even better sensitivities in detecting AGN
candidates than currently available X-ray data below 10 keV for heavily
Compton thick AGNs with column densities of $N_{\rm H} > 10^{24}$
cm$^{-2}$ , although separation of AGN components from star forming
activities could always become an issue \citep{oya11}. To understand 
the efficiency and completeness of these surveys at different wavelengths,
 it is quite important to establish the relation between hard X-rays and infrared
  emission of AGNs based on a large sample of nearby, bright AGNs for which
   detailed studies can be made.

The {\it Swift}/Burst Alert Telescope (BAT) survey \citep{tue08} is one
of the most sensitive all sky surveys in the hard X-ray band ($>$10 keV),
providing us with the least biased sample of AGNs in the local universe
including heavily obscured ones, along with those by INTEGRAL \citep{win03, bir10}.
 {\it Suzaku} follow-up observations of BAT AGNs have discovered a
new type of deeply buried AGNs with a very small scattering fraction
\citep{ued07, egu09, win09a}. Assuming that the amount of
gas responsible for scattering is not much different from other objects,
it is suggested that these new type AGNs are obscured in a geometrically
thick torus with a small opening angle. Understanding the nature of this
population is important to reveal their roles in the cosmological
evolution of SMBHs and host galaxies.

The MIR band also provides crucial information on the inner region
of the AGN tori. It is known that a thermal continuum in the MIR band
 originates from hot circumnuclear dust heated by optical/UV/X-ray photons
from the central engine. Many works suggest that MIR emission is a good
indicator of AGN activity. \cite{hor08,gan09} have found a strong correlation
between X-ray (2--10 keV) and MIR (12.3 $\mu$m) luminosity from the
nucleus of Seyfert galaxies, using the VLT/VISIR data where the AGN can
be spatially resolved from the host galaxy in many cases.
In this paper, we statistically examine the correlation between the
infrared and X-ray luminosities of AGNs by utilizing a large uniform
sample in the local universe, and investigate their infrared properties
as a function of obscuration type. For this purpose, we use the {\it
Swift}/BAT 9-month catalog \citep{tue08} as the parent sample, whose
multiwavelength properties have been intensively investigated. Here we
focus only on ``non-blazar'' AGNs. As for the infrared data, we
primarily use the all-sky survey catalogs obtained with {\it AKARI},
Japanese first infrared astronomical satellite launched on 2006
February 22 \citep{mur07}, which provide unbiased galaxy samples selected in the
mid- and far-infrared (FIR) bands with unprecedented sensitivities as an
all sky survey mission. To complement the infrared data of AGNs whose
counterparts are not detected or do not have reliable flux measurements
with {\it AKARI}, we also utilize the all sky survey catalog of NASA's
Wide-field Infrared Survey Explorer \citep[WISE;][]{wri10} mission,
launched in 2010, as well as the catalogs of the Infrared Astronomical
Satellite \citep[IRAS;][]{neu84}, a joint project of the US, UK, and the
Netherlands launched on January 25, 1983. In Section~2, we present the sample selection
criteria and the results of cross correlation between the Swift/BAT and
AKARI catalogs. In Section 3, we discuss our two main results, the
luminosity correlations and infrared average SED for different types.
The conclusion and summary are given in section 4. Throughout the paper,
we adopt $H_{0} = 70.0$ km s$^{-1}$ Mpc$^{-1}$, $\Omega_{\rm M}=0.3$,
and $\Omega_{\Lambda}=0.7$.

\section{Sample}

\subsection{{\it Swift}/BAT Hard X-ray Catalog}

The Swift/BAT 9-month catalog \citep{tue08} contains 137 non-blazar
AGNs with a flux limit of $2\times 10^{-11}$ erg cm$^{-2}$ s$^{-1}$ in
the 14--195 keV band. 
The redshift range of this sample is $0<z<0.156$.  \cite{win09a}
investigated the soft X-ray (0.5--10 keV) properties of 128 (94.8\%)
BAT-detected non-blazar AGNs of \cite{tue08}. By fitting the X-ray
spectra taken with {\it Swift}/X-Ray Telescope (XRT) or {\it
XMM-Newton}, they derive key spectral parameters, such as the
absorption column density ($N_{\rm H}$), covering fraction of the
absorber ($f_{\rm c}$) or the scattering fraction ($f_{\rm s}$) with
respect to the transmitted component ($f_{\rm scat} \simeq 1-f_{\rm c}$).
In our paper, we do not use the interacting galaxies NGC 6921 and MCG
+04-48-002, which are not separated in the Swift/BAT catalog. Hence,
the parent sample consists of 135 sources.

\subsection{Infrared Catalogs}

\subsubsection{{\it AKARI} Point Source Catalogs}

To obtain the infrared band properties of these Swift/BAT AGNs, we
mainly use the {\it AKARI} All-Sky Survey Point Source Catalogs
(AKARI-PSC). {\it AKARI} carries two instruments, the infrared camera
\citep[IRC;][]{ona07} for the 2--26 $\mu$m band (centered at 9 $\mu$m
and 18 $\mu$m) and the Far-Infrared Surveyor \citep[FIS;][]{kaw07} for
the 50--200 $\mu$m band (centered at 65, 90, 140, and 160 $\mu$m). One
of the major objective of AKARI satellite is to obtain an all-sky map
of infrared sources. The AKARI all-sky survey observations covered
nearly the full sky ($\geq $96 \%), and detected 870,973 sources with
the IRC and 427,071 sources with the FIS. It achieved the flux
sensitivities of 0.05, 0.09, 2.4, 0.55, 1.4, and 6.3 Jy with position accuracies of
6 arcsec at the 9, 18, 65, 90, 140, and 160 $\mu$m bands,
respectively.  In our study, we only utilize sources with the quality
flag of $FQUAL = 3$, whose flux measurements are reliable\footnote{
See the release note of the AKARI/FIS catalog for the details of {\it FQUAL}.
It is recommended not to use the flux data when FQUAL $\leq$ 2 for a reliable scientific analysis.\\
 \url{http://irsa.ipac.caltech.edu/data/AKARI/documentation/AKARI-FIS\_BSC\_V1\_RN.pdf}}. 
As for the FIS catalog, we only refer to the 90 $\mu$m data as a
representative FIR flux, which achieve the most significant
sensitivity improvement compared with the previous {\it IRAS} mission
among the four FIR bands.

\subsubsection{IRAS Catalogs}

The IRAS mission performed an unbiased all sky survey at the 12, 25,
60 and 100 $\mu$m bands. The typical position accuracy at 12
and 25 $\mu$m is 7 arcsec and 35 arcsec in the scan and cross scan
direction, respectively \citep{bei88}. In this paper we use two largest catalogs,
the IRAS Point Source Catalog (IRAS-PSC) and the IRAS Faint Source
Catalog (IRAS-FSC). IRAS achieved 10$\sigma$ point source
sensitivities better than 0.7 Jy over the whole sky. The
IRAS-FSC  contains even fainter sources with fluxes of $>$0.2
Jy in the 12 and 25 $\mu$m bands. 
We use only IRAS sources with {\it FQUAL}$=3$ (the highest quality)
\footnote{
see \cite{bei88} for the definition of {\it FQUAL} in the IRAS catalogs.
False detections may be included when {\it FQUAL} $\leq  2$. }.

\subsubsection{WISE All-Sky Catalog}
The WISE mission mapped the sky at the 3.4, 4.6, 12, and 22 $\mu$m
bands, achieving 5$\sigma$ point source sensitivity better than
0.08, 0.11, 1, and 6 mJy, respectively, in unconfused regions on the
ecliptic poles  \citep{wri10}. 
The WISE all-sky survey 
utilizes the data taken from 2010 January 7 to August 6 \footnote{see the
release note of WISE,\\
http://wise2.ipac.caltech.edu/docs/release/allsky/}.
The source catalog contains positional and photometric information for
over 563 million objects. The position accuracy estimated from the comparison
with the 2MASS catalog is $\sim$2 arcsec at $3\sigma$ level.
In this paper, we only use sources with the flux quality
indicator $ph\_qual=A$, which has Signal-to-Noise ratio larger than 10.
Since the angular resolutions of WISE (6.5 and 12.0 arcsec at 12 and 22 $\mu$m, respectively)
 are slightly worse than those of AKARI IRC (5.5 and 5.7 arcsec at 9 and 18 $\mu$m, respectively), we refer to the
profile-fitting photometry of WISE derived by assuming point-like
sources, for consistency with the AKARI catalog. This is justified
because our targets of WISE are relatively distant (hence more compact)
compared with those detected with AKARI or IRAS (see Section 2.3 for
details).
The photometric data of WISE are given in Vega magnitude, from which we
convert into the Jansky unit using the zero-point flux densities of
$F_{\nu}({\rm iso}) = $ 31.674 Jy and 8.363 Jy for 12 $\mu$m and 22
$\mu$m, respectively.

\subsection{Cross Correlation of Swift/BAT AGNs with the IR Catalogs}

We determine the IR counterparts of the Swift/BAT AGNs by
cross-correlating the AKARI, IRAS, and WISE catalogs in this order.
Our primary goal is to obtain the photometric data in the MIR band as
completely as possible from the hard X-ray selected sample. We put the
highest priority to the AKARI catalog because of its high sky coverage
(97 \% of the all sky) and 2--4 times higher sensitivity than the IRAS
survey. While all the IRAS sources should be detected with AKARI,
AKARI's flux quality flags of very nearby ($z<0.005$) objects turn out
to be bad due to their extended morphology when fitted with a single
Gaussian. In such cases, we rather refer to the IRAS data with good
flux quality, which have $\approx$11 times worse angular resolution
than AKARI, since we aim to measure the total MIR flux from both
nucleus and host galaxy in a uniform way for all the AGN sample. For
AGNs that are not detected with AKARI or IRAS, we utilize the WISE
all-sky catalog, which has 50 times better sensitivity than AKARI and therefore
we can search fainter sources than ever in the MIR all-sky view,
although the bright source are saturated due to the high sensitivity.

\begin{figure}[tbp]
\begin{center}
\includegraphics[angle=0,scale=0.8]{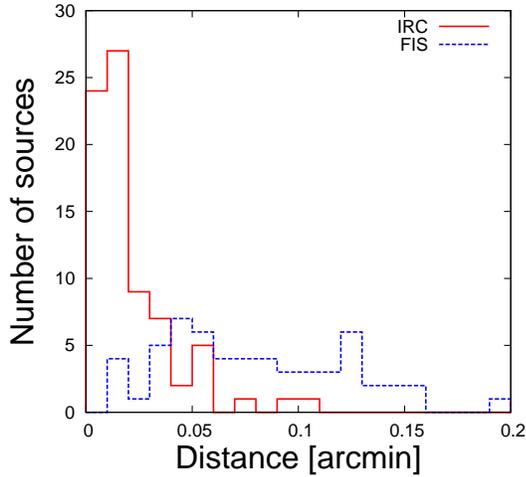}
\\
\caption{
The histograms of position difference between the optical counterparts of the
 Swift/BAT AGNs and their AKARI counterparts (red solid line: IRC, blue dashed line: FIS).
\label{fig-1}}
\end{center}
\end{figure}

First, based on the positional matching of the optical counterparts of
the {\it Swift}/BAT AGNs with the AKARI-PSC, we determine their
infrared counterparts in the 9 $\mu$m, 18 $\mu$m, and 90 $\mu$m
bands. Here we adopt the maximum angular separation of $0.15'$ and
$0.2'$ for the IRC and FIS sources, respectively, which correspond to
typical 3$\sigma$ positional errors at faintest fluxes \citep{ish10, yam10}. We
find 70, 79, and 62 AKARI counterparts in the 9 $\mu$m,
18$\mu$m, and 90 $\mu$m bands out of the total 135 non-blazar BAT
AGN sample. Figure~\ref{fig-1} shows the distribution of the angular
separation between AKARI and optical positions for the Swift/BAT AGNs
with IRC counterparts (red) and those with FIS counterparts (blue). The
IRC sources are more concentrated in a small distance range (with an
average of $\langle \Delta r \rangle = 0.02'$) than the FIS sources
($\langle \Delta r \rangle = 0.08'$), as expected from the positional
accuracy in these catalogs.

Further, for AGNs whose MIR fluxes are not reliably measured
 ($FQUAL < 3$) or not detected with AKARI (65 and 56 sources 
in the 9 $\mu$m and 18 $\mu$m), we search for their counterparts
 at 12 $\mu$m or 25 $\mu$m in the IRAS-FSC and IRAS-PSC.
Here we adopt the 50 arcsec radius, corresponding to
the $<$2$\sigma$ positional error in the cross-scan direction.  As a
result, 11 and 9 IRAS counterparts with $FQUAL=3$ are identified in
the 12 $\mu$m and 25 $\mu$m band, respectively. Finally, we utilize
the the WISE catalog to find the MIR counterparts in the
12 $\mu$m or 22 $\mu$m band for the remaining AGNs detected neither
in the AKARI nor IRAS catalogs (54 and 47 sources in the 9 $\mu$m 
and 18 $\mu$m bands). The matching radius of 2 arcsec is adopted. 
Thus, we identify 45 and 39 WISE sources with 
$ph\_qual=A$ in
the 12 $\mu$m and 22 $\mu$m band, respectively. In summary, we
identify total 128 MIR counterparts detected any in the 9, 12,
18, 22, and 25 $\mu$m out of the 135 Swift/BAT AGNs. Thus, the
completeness of identification in the MIR band is 95\%.

We confirm that the probability of wrong identification with
unassociated IR sources is negligible with these criteria for all the
IR catalogs. Since the mean number density of the AKARI-PSC IRC and
FIS sources in the all sky is $\sim$20 deg$^{-2}$ and $\sim$10 deg$^{-2}$, the
expected number of contamination for the total 135 AGNs within each
error circle is estimated to be only 0.05 and 0.04, respectively.
The number density in the IRAS catalog is 6.2 deg$^{-2}$, and hence the
expected contamination for the 65 AGNs whose counterparts are searched
for within the radius of 50 arcsec is 0.24. Similarly, we estimate
false identification of the WISE sources
with a number density of $ 5.63 \times 10^8 / (4.125\times10^4) =1.36\times10^4$ deg$^{-2}$ to be 0.71
within the radius of 2 arcsec for the searched 54 sources.

\subsection{AGN Type}
To examine the infrared properties for different AGN populations, we
divide the sample into three types based on the X-ray spectra. The first
one is ``X-ray type-1'' (hereafter type-1) AGNs, defined as those
showing the absorption column density of $N_{\rm H} <
10^{22}$cm$^{-2}$. The second is ``X-ray type-2'' (hereafter type-2)
AGNs that have $N_{\rm H} > 10^{22}$cm$^{-2}$. In addition, we are
interested in whether or not there is distinction in the IR properties of
``new type'' AGNs, which exhibit extremely small scattered fraction
($f_{\rm scat} \equiv 1-f_{\rm c}$) suggesting the geometrically thick
tori around the nuclei. Here we define new type AGNs as those satisfying
$f_{\rm c} \ge 0.995$, which are treated separately from the other (normal) 
type-2 AGNs in this paper. 

\subsection{Luminosity Correlation between AKARI, IRAS, and WISE Data}

\begin{figure*}
\begin{center}
\includegraphics[angle=0,scale=.7]{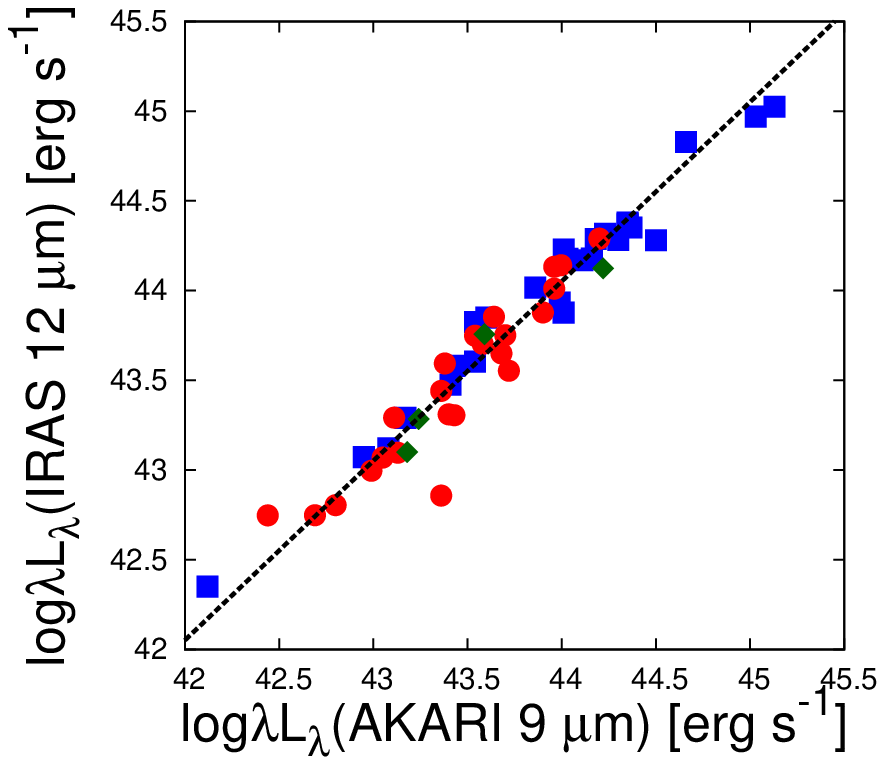}
\includegraphics[angle=0,scale=.7]{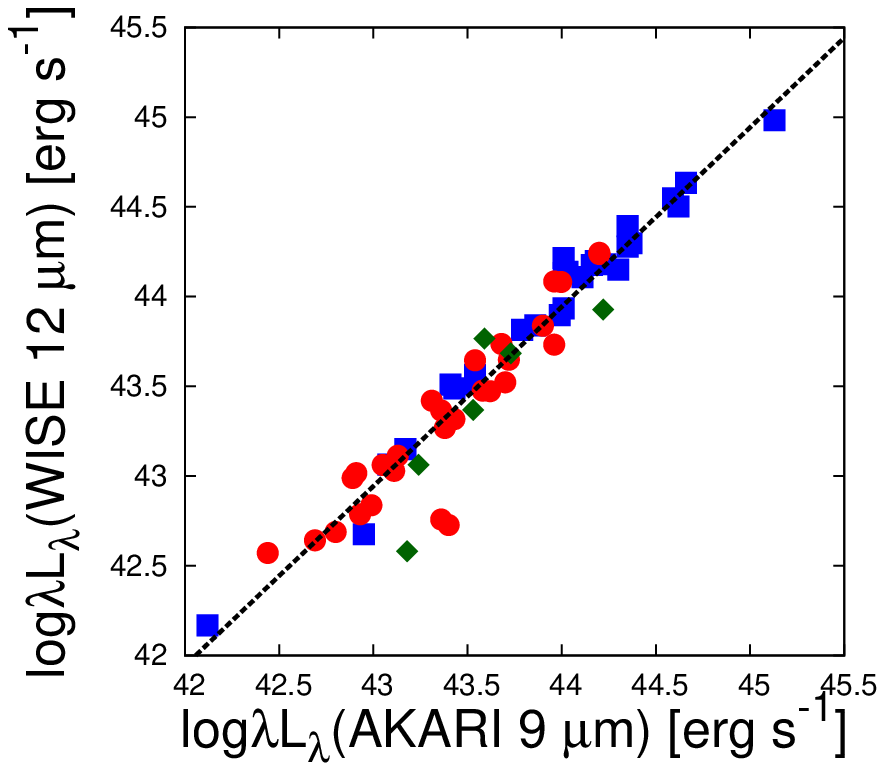}
\includegraphics[angle=0,scale=.7]{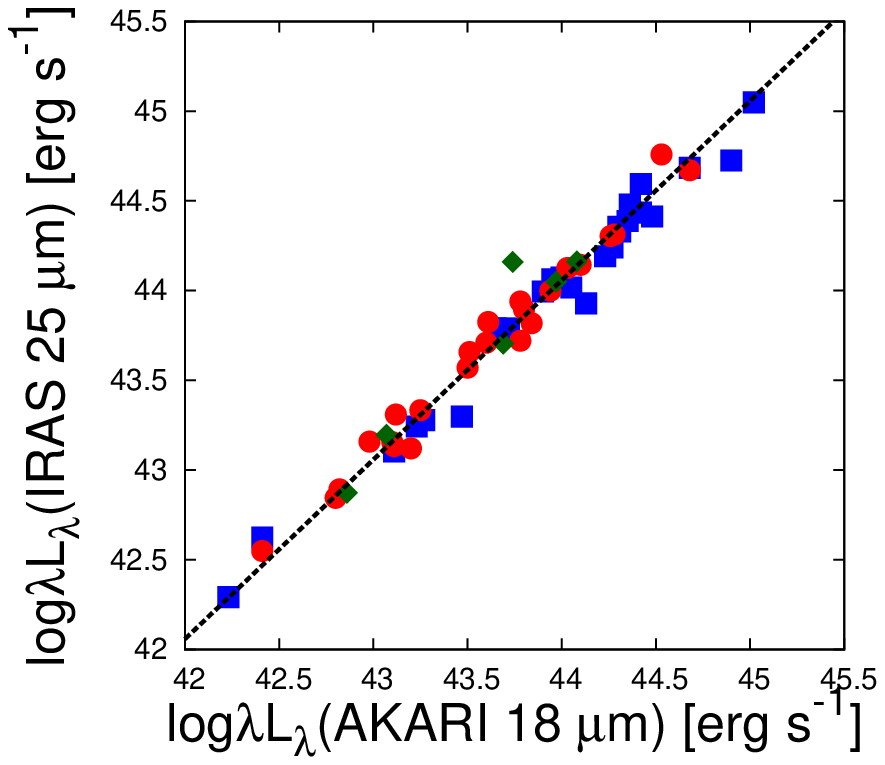}
\includegraphics[angle=0,scale=.7 ]{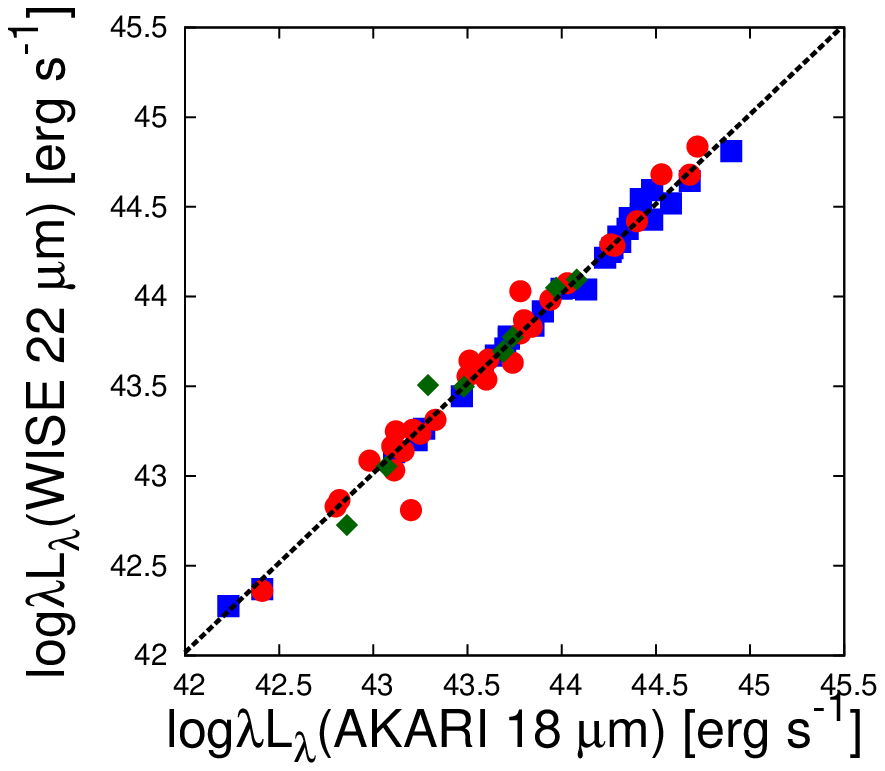}
\\
\caption{
Correlation plots of infrared luminosities between
AKARI 9 $\mu$m and IRAS 12 $\mu$m (top left, 53 sample), 
AKARI 9 $\mu$m and WISE 12 $\mu$m (top right, 61 sample), 
AKARI 18 $\mu$m and IRAS 25 $\mu$m (bottom left, 56 sample), and 
AKARI 18 $\mu$m and WISE 22 $\mu$m (bottom right, 70 sample).
Squares (blue) represent type-1 AGNs ($N_{\rm H} < 10^{22}$cm$^{-2}$), 
circles (red) type-2 AGNs ($N_{\rm H} \geq10^{22}$cm$^{-2}$), 
and diamonds (green) new type AGNs.
The regression lines are given by eqs.\ (1), (2), (3), and (4) in Section~2.5.
\label{fig-2}}
\end{center}
\end{figure*}

To make it possible to uniformly treat the MIR luminosities of AGNs at
slightly different wavelengths obtained from the three IR
observatories, we here investigate the correlation between the
AKARI/IRAS/WISE luminosities, using the sample commonly detected with
AKARI and IRAS, or with AKARI and WISE. 
We choose IRAS 12 $\mu$m/WISE 12 $\mu$m for AKARI 9
$\mu$m, and IRAS 25 $\mu$m/WISE 22 $\mu$m for AKARI 18 $\mu$m, respectively,
because of the proximity of the central wavelengths.
Figure~\ref{fig-2} displays
luminosity correlations between (1) AKARI 9 $\mu$m versus IRAS 12
$\mu$m, (2) AKARI 9 $\mu$m versus WISE 12 $\mu$m, (3) AKARI 18 $\mu$m
versus IRAS 25 $\mu$m, and (4) AKARI 18 $\mu$m versus WISE 22 $\mu$m.
We check the strength of these luminosity correlations 
by using Spearman's test.
We obtain Spearman Rank coefficient ($\rho$) and null hypothesis
probability $P$ of (1) $(\rho, P) =( 0.97, 2.37\times10^{-34})$, (2)
$(\rho, P) = (0.97, 3.6\times10^{-37})$, (3) $(\rho, P) = (0.98,
5.7\times10^{-40})$, and (4) $(\rho, P) = (0.99, \le 10^{-42})$. 
Since correlations between ``luminosities'' can be forced
from those between ``fluxes'', we also check the strength of the flux-flux
correlations and obtain (1) $(\rho, P) =( 0.93, 7.6\times10^{-24})$, (2) $(\rho,
P) = (0.92, 4.9\times10^{-26})$, (3) $(\rho, P) = (0.96,
5.9\times10^{-33})$, and (4) $(\rho, P) = (0.99, \le \times10^{-42})$. 
Thus, we confirm that the correlations in both luminosity and flux
between different infrared catalogs are tight and significant.
The standard deviation of
the luminosity-ratio distribution between these two bands in the
logarithm scale is found to be (1) 0.14 dex, (2) 0.17  dex, (3) 0.10
dex, and (4) 0.08 dex, respectively. The dispersion does not affect our
conclusion on the MIR and hard X-ray luminosity correlation (Section~3.1).
Based on the correlation, we derive the empirical formula to convert the
 IRAS or WISE luminosities at 12 $\mu$m, 22 $\mu$m, or 25 $\mu$m 
 into the equivalent AKARI luminosities at 9 $\mu$m or 18 $\mu$m as follows;
 \begin{align}
\log \lambda L_{\lambda}({\rm AKARI}\ 9\ \mu {\rm m}) =& \log \lambda L_{\lambda} ({\rm IRAS}\ 12\ \mu {\rm m}) -0.051 \\
\log \lambda L_{\lambda }({\rm AKARI}\ 9\ \mu {\rm m}) =& \log \lambda L_{\lambda} ({\rm WISE}\ 12\ \mu {\rm m}) +0.057\\
\log \lambda L_{\lambda}({\rm AKARI}\ 18\ \mu {\rm m}) =& \log \lambda L_{\lambda} ({\rm IRAS}\ 25\ \mu {\rm m}) -0.058 \\
\log \lambda L_{\lambda}({\rm AKARI}\ 18\ \mu {\rm m}) =& \log \lambda L_{\lambda} ({\rm WISE}\ 22\ \mu {\rm m}) -0.016
\end{align}

Assuming that AGNs detected not with AKARI but with IRAS or WISE
should follow the same correlations as examined here, we apply these
conversion factors to derive the 9 or 18 $\mu$m ``AKARI equivalent''
luminosities for them so that we can discuss the correlation with 
hard X-rays in a uniform way regardless of the matched catalogs.
Among the 128 Swift/BAT AGNs with MIR counterparts, 126 and
127 objects have the flux measurement in the 9 $\mu$m and 18 $\mu$m band,
respectively. The 9 $\mu$m sample consists of 70 AKARI
sources, 11 IRAS sources, and 45 WISE sources (126 sources in total),
while the 18 $\mu$m sample consists of 79 AKARI sources, 9 IRAS
sources, and 39 WISE sources (127 sources in total).  The summed sample
detected either in the 9 $\mu$m, 18 $\mu$m, or AKARI 90 $\mu$m band
consists of 128 sources (9 $\mu$m; 126, 18 $\mu$m; 127, 90 $\mu$m; 62).

\subsection{Basic Properties of the Sample}

\begin{figure}[tbp]
\begin{center}
\includegraphics[angle=0,scale=.4]{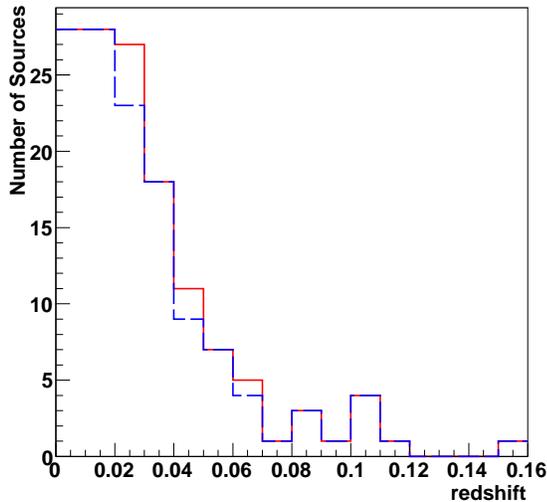}
\caption{
Redshift distribution of the whole non-blazer AGNs in the Swift/BAT
 9-month catalog (red solid line: 135 objects) and of those with the MIR
 counterparts (blue dashed line: 128 objects).
\label{fig-3}}
\end{center}
\end{figure}

Figure~\ref{fig-3} displays the redshift distribution of the 128 AGNs
with MIR counterparts (blue) together with that of the whole
\citep{tue08} sample of 135 AGNs (red). 
Since our sample is highly complete(95\%),
 we regard our current sample with MIR counterparts as
representative of the whole population of hard X-ray selected AGNs, and
hereafter ignore any issues related to the incompleteness.
Table~\ref{tbl-1} summarizes the infrared to X-ray properties of all the
135 Swift/BAT 9 month non-blazar AGNs in \cite{tue08}, including 128 objects
with MIR counterparts: (1) source No.\ in \cite{tue08}, (2) object name, (3)
redshift, (4)--(6) infrared fluxes ($F_\nu$) at 9 $\mu$m, 18
$\mu$m, and 90 $\mu$m, (7)--(9) infrared luminosities ($\lambda
L_\lambda$) at 9 $\mu$m, 18 $\mu$m, and 90 $\mu$m, (10) reference
catalog for the IR data for 9 $\mu$m, 18 $\mu$m, and 90 $\mu$m, (11)
hard X-ray flux in the 14--195 keV band, (12) hard X-ray luminosity in
the 14--195 keV band ($L_{\rm HX}$), (13) X-ray absorption column density ($N_{\rm
H}$), (14) covering fraction in the X-ray spectrum ($f_{\rm c}$), and
(15) the reference for the X-ray spectra.  For AGNs whose AKARI MIR
fluxes are not available, we convert the infrared fluxes and luminosities with IRAS or
WISE into those at 9 $\mu$m or 18 $\mu$m according to the formula given
in Section~2.5. Columns (1), (2), (3), (11) are taken from
\cite{tue08}\footnote{The redshifts of two AGNs,  No. 13: 2MASX 
J02162987+5126246  and No. 53: 2MASX J06403799--4321211,
 are adopted from \url{http://heasarc.gsfc.nasa.gov/docs/swift/results/bs58mon/}}
. The X-ray spectral information (columns 13--14)
is basically adopted from \cite{win09a}, while we refer to the results
obtained with {\it Suzaku}  \citep{egu09, egu11, ris09, tur09, awa08, gon11, win09b, ito08, taz11, win10, bia09}
and those with {\it XMM-Newton} \citep{nog10, bal05} whenever available. 
When the information is not available in \cite{win09a}, we refer to \cite{tue08}.
  All luminosities in this table are calculated by using the redshift given in column (3).
There are total 13 new type AGNs out of the 135 AGNs, for which asterisks are
 attached to the source No.\ in Table~1 (column 1).

\begin{figure}
\begin{center}
\includegraphics[angle=0,scale=0.4]{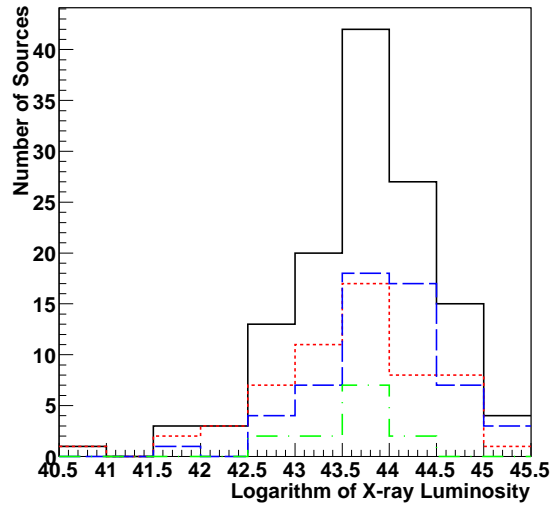}
\caption{
Distribution of the hard X-ray (14--195 keV) luminosity of our sample
with MIR counterparts (total: black). The dashed blue, dotted red, and dot-dashed green ones
correspond to those of the type-1, type-2, and new type AGNs, respectively.
\label{fig-4}}
\end{center}
\end{figure}

Figure~\ref{fig-4} plots the hard X-ray luminosity distribution of our AGN sample
with the MIR counterparts. Those for type-1 (dashed blue), type-2 (dotted red), and
new-type (dot-dashed green) are separately plotted.
Previous studies on Swift/BAT selected AGNs \citep{win09a, bur11} 
have shown that the X-ray luminosity distribution of type-1 AGNs 
has a higher peak luminosity than that of type-2 ones, 
as expected from the well-known correlation that the fraction of
absorbed AGNs decreases against luminosity \citep[e.g.,][]{ued03}.
Although such trend may not be clear in Figure~\ref{fig-4} due to the
smaller sample size and coarse bin width (0.5 dex),
the averaged logarithmic luminosity of the type-1 and type-2 AGNs in our
sample is 43.9 and 43.6, respectively, and 
a K-S test applied to their distributions returns a null probability of 
$4.0 \times 10^{-2}$. Thus, albeit marginal,
type-1 AGNs are more luminous on average than type-2 AGNs in our sample.

\section{Results and Discussion}

\subsection{Correlation between the Infrared and Hard X-ray Luminosities}

\begin{figure*}
\includegraphics[angle=0,scale=.7]{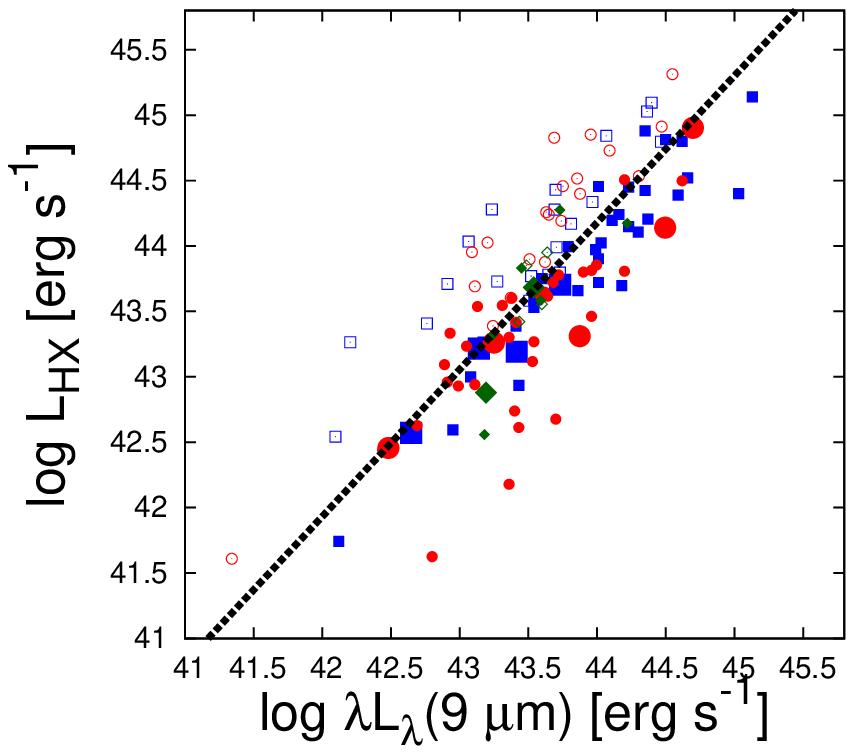} 
\includegraphics[angle=0,scale=.7]{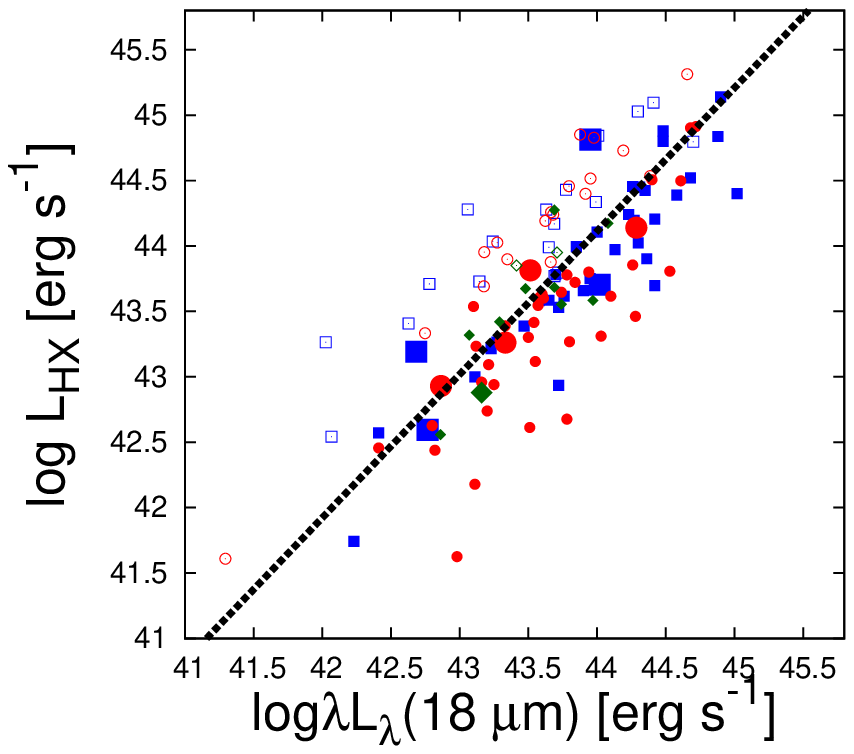}
\includegraphics[angle=0,scale=.7]{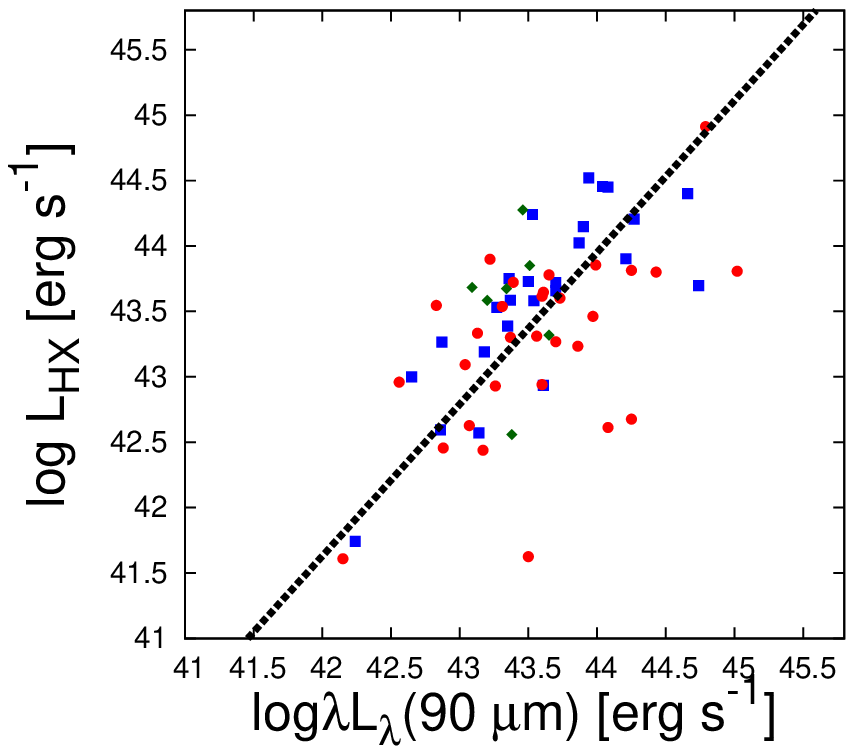} \\
\caption{
Correlation between the infrared (9 $\mu$m, 18 $\mu$m, or 90 $\mu$m) 
and hard X-ray (14--195 keV) luminosities.
 Squares (blue) represent type-1 AGNs ($N_{\rm H} < 10^{22}$cm$^{-2}$), 
 circles (red) type-2 AGNs ($N_{\rm H} \geq10^{22}$cm$^{-2}$),
  and diamonds (green) new type AGNs. Dotted lines represent
 the regression lines (see Section~3.1). 
The small-filled, large-filled, small-open symbols denote
those with the AKARI, IRAS, and WISE counterparts, respectively. 
The IRAS and WISE fluxes are all converted into the 9 $\mu$m (left) or
 18 $\mu$m (right) using the formula given in Section~2.5.}
\label{fig-5}
\end{figure*}

Figure~\ref{fig-5} shows the luminosity correlations in the luminosity
range from $10^{41}$ erg s$^{-1}$ to $10^{46}$ erg s$^{-1}$  between the
infrared (9, 18, or 90 $\mu$m) and {\it Swift}/BAT hard X-ray bands. Type-1,
type-2, and new type AGNs are marked with squares (blue), circles (red),
and diamonds (green), respectively. The small-filled symbols denote the
data from AKARI, large-filled ones those from IRAS, and small-open ones
those from WISE.  In the figure, NGC 4395 is not shown
due to its low luminosities ($\log \lambda L_{\lambda} (9~\mu {\rm m}), 
\log \lambda L_{\lambda} (18~\mu{\rm m}), \log L_{{\rm HX})} = (39.98, 40.28, 40.81)$. 
As noticed from the figure, the MIR (both 9 and 18
$\mu$m) luminosities well correlate with hard X-ray luminosity over 3
orders of magnitude (from 10$^{42}$-- 10$^{45}$ erg s$^{-1}$). In the
FIR (90 $\mu$m) band, by contrast, the correlation is much weaker with
larger dispersion compared with the MIR bands, even though we plot here
only for AGNs detected with AKARI at 90 $\mu$m.

Least-square fits to the hard X-ray versus MIR luminosity plots
with a power law model (i.e., a linear function for the logarithmic luminosities with
 the form of $\log(L_{\rm HX}/10^{43}) = a + b \log(\lambda L_{\lambda}(9, 18~\mu {\rm m})/10^{43})$ give the following best-fit correlations:
\begin{align}
\log \frac{L_{\rm HX}}{10^{43}} = (0.06\pm 0.07) +(1.12 \pm 0.08) \log \frac{\lambda L_{\lambda}(9~ \mu{\rm m}) } {10^{43}}\\
\log \frac{L_{\rm HX}}{10^{43}} = (0.02 \pm 0.07)+(1.10 \pm 0.07) \log \frac{\lambda L_{\lambda}(18~ \mu{\rm m}) } {10^{43}}
\end{align}

To check the significance of the
correlations between the hard X-ray and MIR/FIR luminosities (or fluxes),
we perform Spearman's tests for the summed sample
consisting of all AGN types. The results are summarized in 
Table~\ref{tbl-2}, which has the following columns:
Col. (1) sample; 
Col. (2) number of objects; 
Col. (3) luminosity-luminosity correlation coefficient ($\rho_L$);
Col. (4) flux-flux correlation coefficient ($\rho_f$);
Col. (5) a standard Student's $t$-test null significance level for luminosity-luminosity
correlations ($P_L$);
Col. (6) a standard Student's $t$-test null significance level for flux-flux
correlations ($P_f$);
Col. (7) regression intercept (a) and its 1-$\sigma$ uncertainty;
Col. (8) regression slope (b) and its 1-$\sigma$ uncertainty.
We find that both luminosity-luminosity and flux-flux correlations 
between the hard X-ray and MIR bands are highly significant. 
While there is also a significant correlation between 90 $\mu$m and hard
X-ray luminosity, $(\rho_L, P_L)=(0.59,4.5\times10^{-7})$,
their flux-flux correlation is weak with $(\rho_f, P_f)=(0.17, 0.18)$.

We establish the good correlation between the MIR and hard X-ray
luminosities in AGNs from so far the largest, uniform AGN sample in the
local universe, although similar results have been reported by several
authors \citep{mus08, gan09, vas10}. The MIR emission from galaxies hosting an AGN is
believed to originate mainly from high temperature ($\sim$ 150--300 K)
dust emission heated by X-ray/UV photons from the central engine. Thus,
if extinction is not important, the MIR luminosity is expected to be
correlated with the intrinsic X-ray luminosity, the most reliable
tracers of the AGN power (since our sample contains
mostly Compton thin AGNs with $N_{\rm H} < 10^{24}$
cm$^{-2}$, the observed 14--195 keV luminosity can be regarded as the
intrinsic one without any correction). On the other hand, the FIR
emission comes both from cooler ($\sim 30$ K) interstellar dust heated
by stars in host galaxies and from the cooler (outer) part of the
torus in the AGN. The contribution from the host galaxy increases the
scatter of the observed luminosity correlation, depending on the total
star forming rate over the whole galaxy. Our results demonstrate that
the MIR emission is more suitable for estimating the AGN intrinsic
power than in the FIR band.

\begin{figure*}
\begin{center}
\includegraphics[angle=0,scale=.8]{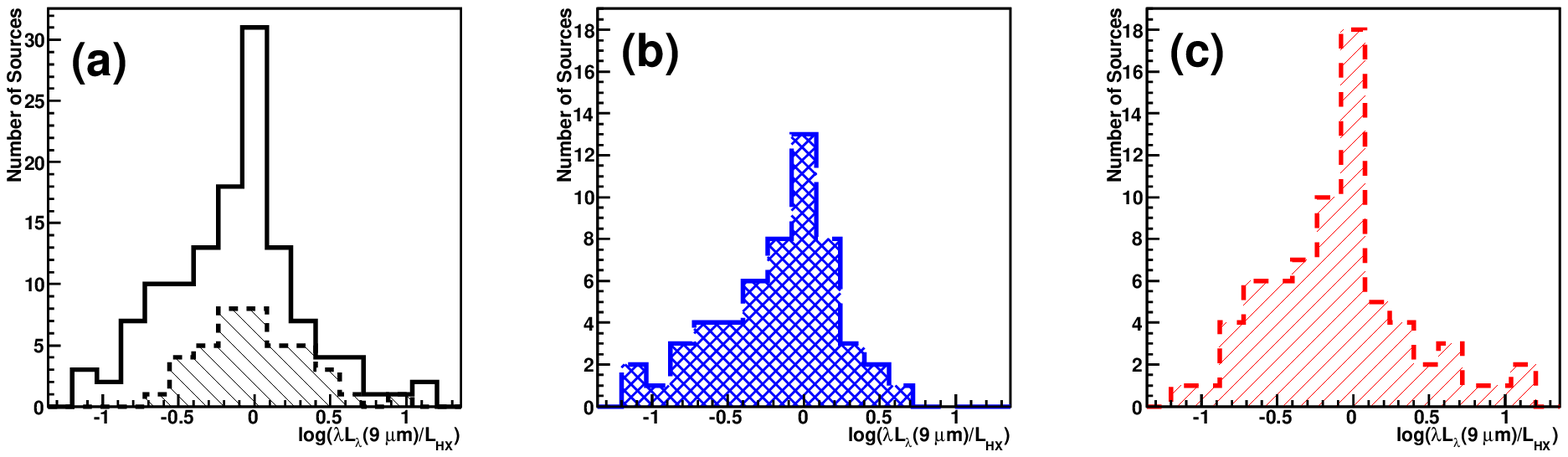}
\includegraphics[angle=0,scale=.8]{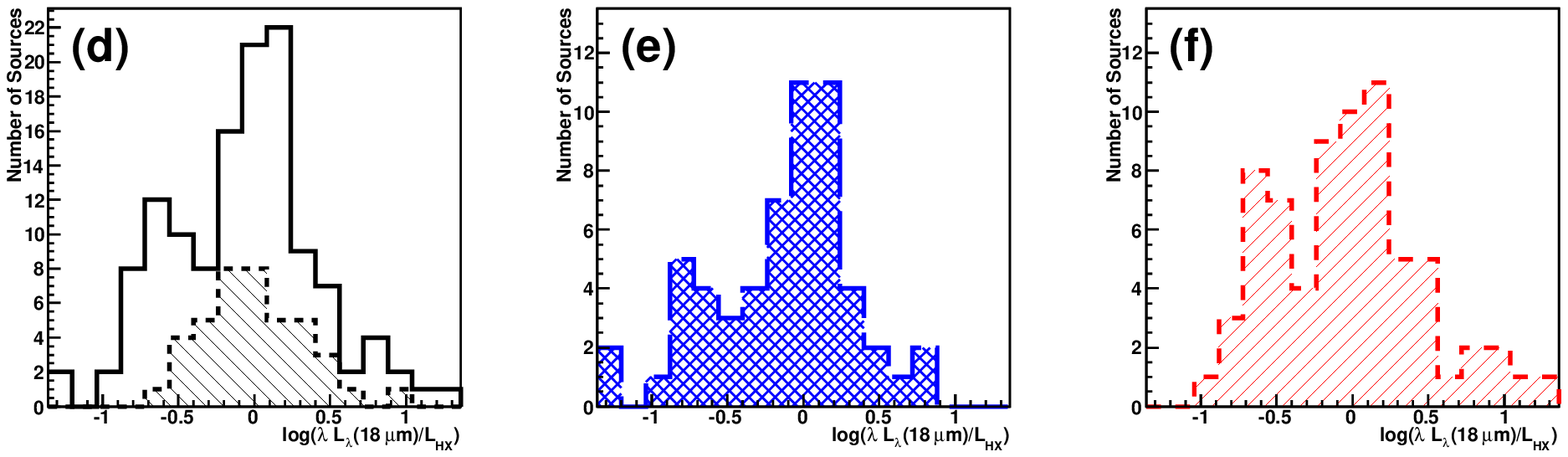}
\\
\caption{
Histograms of $\log (\lambda L_{\lambda}(9~\mu {\rm m})/L_{\rm HX})$ (top) and
$\log (\lambda L_{\lambda}(18~\mu {\rm m})/L_{\rm HX})$ (bottom).
(a) and (d) (left): total AGNs (solid lines),
(b) and (e) (center): type-1 AGNs, 
(c) and (f)  (right): type-2 + new type AGNs
That of $\log (\lambda L_{\lambda}(12.3~\mu {\rm m})/L_{\rm HX})$ for the 
\cite{gan09} sample is overplotted in panel (a) and (d) (dashed lines, hatched area).}
\label{fig-6-1}
\end{center}
\end{figure*}

Figures~\ref{fig-6-1}
plot the histograms of the MIR to hard X-ray luminosity ratio in the
logarithm scale ($r \equiv \log \lambda L_{\lambda}(9~\mu {\rm m})/L_{\rm HX}$ and $\log
\lambda L_{\lambda}(18~\mu {\rm m})/L_{\rm HX}$) for the Swift/BAT AGN sample with MIR
counterparts in the 9 $\mu$m and 18 $\mu$m bands, respectively,
calculated from (a, d) total, (b, e) type-1 AGNs, and (c, f) type-2 + new type
AGNs.  Following \cite{gan09}, we calculate its average and standard
deviation for each sample, which are summarized in Table~3. From the
total AGNs, we obtain $(\bar{r}, \sigma)$ = $(-0.129\pm0.039, 0.437 \pm0.055)$ for
the 9 $\mu$m band and $(\bar{r}, \sigma)$ = $(-0.080\pm0.042, 0.473\pm0.059)$ for
the 18 $\mu$m band. We find that the average is consistent between
type-1 and type-2 (plus new type) AGNs within the errors. 

For comparison, the histogram of $\log \lambda L_{\lambda}(12.3~\mu {\rm
m})/L_{\rm HX}$ obtained by \cite{gan09} from 42 local AGNs are
overplotted in Figures~\ref{fig-6-1}(a) and (d), where we
convert their 2--10 keV luminosities into the 14--195 keV band by
assuming a power law photon index of 2. Their sample is
selected from nearby AGNs with good available X-ray spectra observed
with {\it Suzaku}, {\it INTEGRAL}, or {\it Swift}, whose average
distance is 74 Mpc (or $z=0.017$), consisting of 12 Seyfert 1s, 19
Seyfert 2s, 3 LINERs, and 8 Compton thick ($N_{\rm H} \ge 1.5 \times
10^{24}$cm$^{-2}$) AGNs. It is not a statistically complete sample,
however. The average X-ray luminosity is $\sim 42.9$ and only 3 sources
have QSO class luminosity ($L(2-10~{\rm keV}) > 10^{44}$erg s$^{-1}$). Twenty
four objects out of the 42 AGNs are listed in the Swift/BAT 9-month
catalog and in our Table~1.  We obtain $(\bar{r}, \sigma)$ =
$(0.086\pm0.054, 0.353\pm0.077)$ for the 12.3 $\mu$m band from the \cite{gan09}
sample. Although the average is slightly larger than ours, it is still
consistent within the error in the 18 $\mu$m band. Note that silicate
features (see below) may affect our result in the 9 $\mu$m band. The
standard deviation is almost the same between the \cite{gan09} sample
and ours; the small difference could be due to the fact that there are
22 ``well-resolved'' AGNs in \cite{gan09}, whose
nucleus MIR emission was spatially separated from the host galaxy,
making the correlation between the MIR and X-ray luminosities tighter.

We also check the consistency with the results by
\cite{mat12} obtained from the cross correlation between the Swift/BAT
22 month sample and the AKARI catalog. We confirm that our best-fit
slopes ($b$) of the linear correlations between log $\lambda L_{\lambda}(9, 18~\mu {\rm
m})$ and log $L_{\rm HX}$ are well consistent with their results (1.13
for 9 $\mu$m and 1.12 for 18 $\mu$m with average errors of 0.04). On
the other hand, the averaged MIR to X-ray luminosity ratios
($\bar{r}$) derived from our study are slightly smaller than those by
\cite{mat12} (0.14 for 9 $\mu$m and 0.19 for 18 $\mu$m). The
difference is attributable to the high completeness of identification in our analysis,
where faint, WISE-only detected MIR sources are included.


\begin{figure*}
\begin{center}
\includegraphics[angle=0,scale=.9]{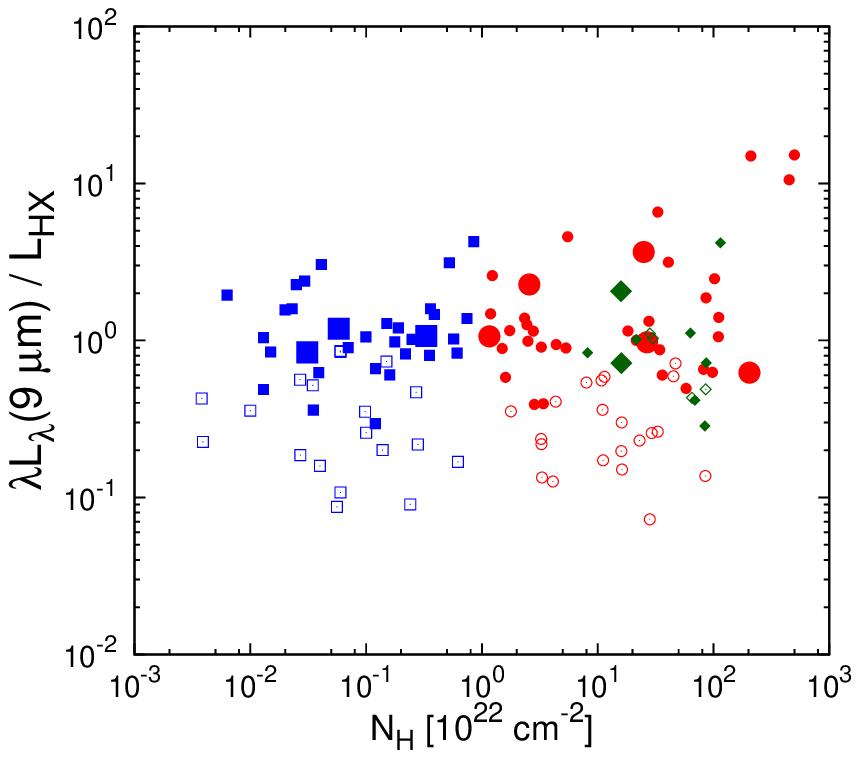}
\includegraphics[angle=0,scale=.9]{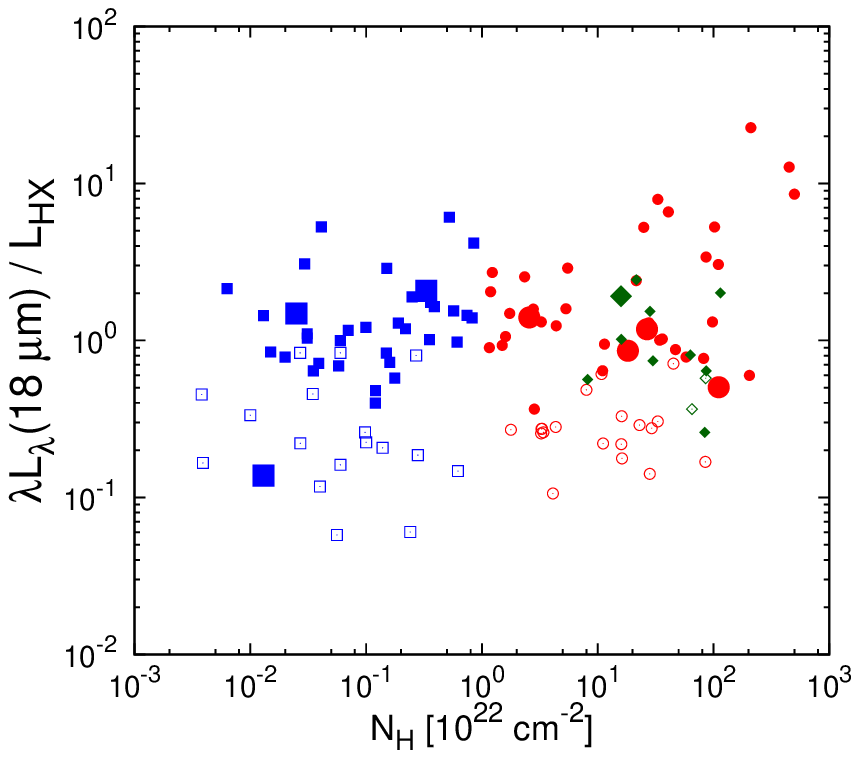}\\
\caption{
The MIR to hard X-ray luminosity ratio plotted against the absorption 
column density. 
Left: $\lambda L_{\lambda}(9~\mu{\rm m})/L_{\rm HX}$. 
Right:$\lambda L_{\lambda}(18~\mu{\rm m})/ L_{\rm HX}$.
All symbols are the same as Figure~\ref{fig-5}.
\label{fig-7}
}
\end{center}
\end{figure*}

As noticed from Figure~\ref{fig-5}, all types of AGNs (type-1, type-2,
and new type) seem to follow almost the same MIR vs hard X-ray
luminosity correlation.  In fact, we see no significant difference in
the distribution of their luminosity ratio between type-1 and
type-2/new type AGNs (Table~3). To check this further, we plot
$\lambda L_{\lambda}(9~\mu {\rm m})/ L_{\rm HX}$ and $\lambda
L_{\lambda}(18~\mu {\rm m})/ L_{\rm HX}$ as a function of the
absorption column density ($N_{\rm H}$) in Figure~\ref{fig-7}. In both
panels, there is no clear dependence of the MIR to X-ray luminosity
ratio on $N_{\rm H}$ up to log $N_{\rm H} \simeq
24$. The large ratios found for Compton thick AGNs can be partially
explained by attenuation of the hard X-ray fluxes due to heavy
obscuration. The absence of $N_{\rm H}$ dependence
suggests that the emission from the AGN-heated dust seems not to
be affected by the obscuration by the torus causing the X-ray
absorption.  The results cannot be explained with homogeneous dust
torus models \citep{pie92, pie93}, which predict the significant
decrease in the MIR to X-ray luminosity ratio when an optically-thick
line-of-sight through the torus primarily show cooler smooth-dust and
a lower mid-infrared luminosity for the same X-ray luminosity than
does an optically-thin one.  Our results rather favor
 the clumpy dust tori model \citep{hon06, nen08a, sch08}
, which predicts isotropic MIR emission, as discussed in \cite{gan09}. No clear
difference is seen in the MIR to X-ray luminosity correlation between
normal AGNs and new type AGNs. New type AGNs may have large intrinsic
MIR luminosities because of the geometrically thick torus
\citep{nen08b}, which could be partially canceled out due to
extinction, however. Currently, the possible contribution from the
host galaxy in the MIR band makes direct comparison between individual
objects very difficult. We need more observations of these AGNs by
resolving only the nucleus emission as done in \cite{gan09}.

\cite{mel08a, mel08b, wea10} proposed the [Ne V]
  14.32/24.32 $\mu$m and [O IV] 25.89 $\mu$m lines as good AGN
  indicators because of their high ionization potential. Using a Swift/BAT 
  sample, \cite{wea10} found a tight luminosity correlation between these
   lines and hard X-ray luminosities with a scatter of $\sim$0.5 dex, which
   is almost comparable with our result obtained for the MIR and hard X-ray
    correlation ($\approx$0.45 dex, see Table~\ref{tbl-2}).
    The correlation we find is very useful and easy to apply to various survey data as it
  requires only photometry without spectroscopy.

\subsection{Averaged IR Spectral Energy Distribution}

Spectra in the MIR band are quite useful to investigate the host galaxy
properties of AGNs.  It is known that Polycyclic Aromatic Hydrocarbon
(PAH) features in the 3 $\mu$m to 20 $\mu$m band \citep{tie08} can be
used as a starburst tracer \citep{ima00}. The AKARI 9 $\mu$m band covers
the strong PAH emission feature at 7.7, 8.6, and 11.2
$\mu$m. In addition, opacity peak of
amorphous silicate grains are located around 10 and 18 $\mu$m due to
the Si$-$O stretching and the O$-$Si$-$Y bending modes.
\cite{hao07} report the MIR spectra of different types of AGNs (type-1
QSO, type-1 Seyfert, type-2 Seyfert, ULIRG) from Spitzer observations,
revealing a large variety in the 10 $\mu$m silicate feature.
Silicate absorption feature is clearly detected from type-2 Seyferts.

\begin{figure}
\begin{center}
\includegraphics[angle=0,scale=0.9]{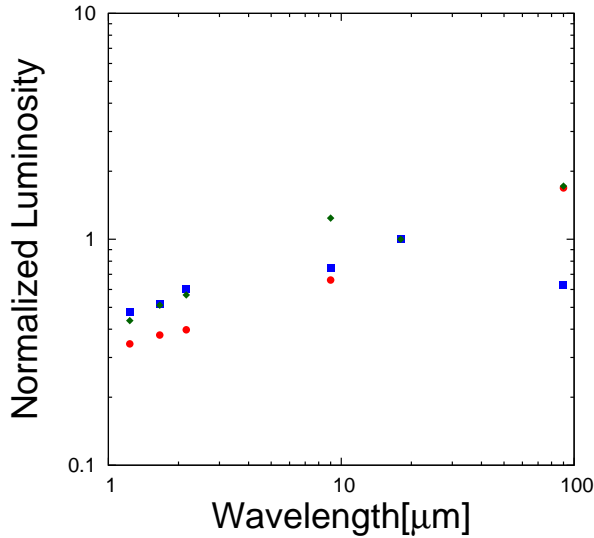} 
\caption{
Averaged infrared (1--100 $\mu$m) SED normalized at the 18 $\mu$m band
for type-1 (blue, filled square), type-2 (red, filled circle), and new type
AGNs (green, diamond).
\label {fig-8}
}
\end{center}
\end{figure}

Averaged SEDs in the NIR to FIR band of the three types of AGNs are
presented in Figure~\ref{fig-8}. The spectrum of each AGN is
normalized by the 18 $\mu$m luminosity. Here we only use total 42
sources detected in all the 9, 18, and 90 $\mu$m bands, consisting of
16 type-1, 21 type-2, and 5 new type AGNs. We also include the
photometric data in the $J$, $H$, and $K_{\rm s}$ bands adopted from
the 2 Micron All Sky Survey (2MASS) Point Source Catalog. We neglect
the effect of redshift because the sample consists of only local AGNs
($z < 0.1$ with a mean value of $\langle z \rangle=0.0165$), for which
K-correction is not significant.

We find that the FIR emission at 90 $\mu$m is weaker relative to 18
$\mu$m in the type-1 AGNs (blue) than in type-2 (red) and new type
AGNs (green). On the other hand, the MIR spectra are almost the same
between the type-1 and type-2 AGNs. This can be explained because
type-1 AGNs have intrinsically higher AGN luminosities on average than
type-2 AGNs (see Figure~\ref{fig-4}), and hence contribution from cool
dust in the host galaxy emitting the FIR radiation becomes smaller
relative to the AGN component mainly observed in the MIR band.
Indeed, the same trend is seen in the observed SED
templates of Seyfert 2 galaxies and type-1 QSOs complied by
\cite{pol07} (see their Figure~1). Another possible
explanation would be an intrinsic difference of the dust quantity
between type-1 and type-2 AGNs: \cite{mal98} suggested that the host
galaxies of type-2 AGNs are likely to be more dusty than those of
type-1 AGNs from the results of a Hubble Space Telescope imaging
survey of nearby AGNs.

It is remarkable that the averaged SED of new type AGNs exhibit
enhanced fluxes than type-2 AGNs at 9 $\mu$m. In fact, this is
confirmed in the individual SED for 4 objects (ESO 005-G004,
 ESO 506-G027, NGC 7172 and NGC 7319) out of the 5 new
type AGNs examined here. This 9 $\mu$m excess most probably
attributable to the PAH emission feature at 7.7, 8.6, and 11.2 $\mu$m
 from the host galaxies. This is consistent with the
larger 90 $\mu$m excess in the averaged new-type AGN spectrum than in
type-2 AGNs, which also reflects the starburst activity in the host
galaxies. Another possibility is that the 9 $\mu$m excess comes from
the emission feature of silicate grains, which would imply a larger
amount of dusts around the nucleus than in normal type-2 AGNs. Our
result suggests that geometrically thick tori around the black hole
may form in galaxies with high star forming rates.

Four out of the 5 new type AGNs have available MIR
spectra observed with {\it Spitzer}. ESO 005-G004 shows a clear 11.2
$\mu$m PAH line, and silicate absorption features are suggested at
$\lambda > 10~\mu$m \citep{wea10}. A detection of the 11.2 $\mu$m PAH
line is reported from ESO 506-G027 \citep{sar11}, although the
spectrum is not available in the literature.
NGC 7172 shows PAH lines at 7.7 $\mu$m and 11.2 $\mu$m
with strong silicate absorption features. In addition, for ESO
005-G004 and NGC 7172, the line flux ratio between [Ne III] 15.56
$\mu$m and [Ne II] 12.81 $\mu$m is available in \cite{wea10}, from
which the relative strength of the starburst to AGN activities can be
estimated. These two sources have $\log$ $f_{\rm [Ne III]}/f_{\rm [Ne
II]}$ = $-0.50$ and $-0.31$, respectively, suggesting that both have
relatively strong starburst components normalized by their AGN
activities. Thus, though limited in sample, a majority of our new type
AGNs indeed show significant PAH emission lines (and silicate
absorption features) contributing to the 9 $\mu$m excess.
Further
systematic investigation of the MIR spectra of hard X-ray selected
AGNs is useful to reveal the host galaxy properties and environment
around the central engine in relation to the X-ray spectral
information.

\section{Summary and Conclusions}

We have systematically studied the MIR and FIR properties of a large
complete flux limited AGN sample in the local universe detected in the
Swift/BAT all sky survey in the 14--195 keV band, which has the least
bias against obscuration. Utilizing the AKARI, IRAS, and WISE infrared
catalogs, we unambiguously identify 128 counterparts in the MIR band
out of the 135 non-blazar AGNs in the Swift/BAT 9-month catalog by
\cite{tue08}. For our discussion, the whole sample is divided into 3
types based on the X-ray spectra, 57 type-1, 58 type-2, and 13 ``new
type'' AGNs showing extremely small scattered fractions.

The two main conclusions are summarized as follows: 
\begin{enumerate}

\item We find a good luminosity correlation between the MIR (9 $\mu$m
      and 18 $\mu$m) and hard X-ray band over three orders of
      magnitude ($42<\log L_{\rm HX} < 45$), while that between the
      FIR (90 $\mu$m) and hard X-ray bands is weaker, most probably
      due to the larger contribution from the host galaxy
      (Figure~\ref{fig-5}). All types of AGNs follow the same
      correlation, and the luminosity ratio between the MIR to X-ray
      bands show no clear dependence against absorption column density
      up to $N_{\rm H} \sim 10^{24}$ cm$^{-2}$. Our results favor
      isotropic infrared emission models, possibly clumpy dust torus
       models rather than homogeneous dust model,
      confirming the argument by \cite{gan09} but with a much larger
      sample.
     
\item We find 9 $\mu$m excess in the averaged infrared SED of ``new type''
	AGNs. This could be attributable to the PAH emission features,
	as confirmed in the available {\it Spitzer}
	spectra of at least three sources, suggesting that their
	host galaxies have strong starburst activities. 

 \end{enumerate}

\acknowledgments

This research is based on observations with AKARI, a JAXA project with
the participation of ESA, and those with Wide-field Infrared Survey
Explorer, which is a joint project of the University of California,
Los Angeles, and the Jet Propulsion Laboratory/California Institute of
Technology, funded by the National Aeronautics and Space
Administration. The publication makes use of the NASA/
IPAC Infrared Science Archive, which is operated by the Jet Propulsion
Laboratory, California Institute of Technology, under contract with
the National Aeronautics and Space Administration. This work is
partly supported by the Grant-in-Aid for Scientific Research 23540265
(YU) and 21244017 (YT), and by the grant-in-aid for the Global COE
Program ``The Next Generation of Physics, Spun from Universality and
Emergence'' from the Ministry of Education, Culture, Sports, Science
and Technology (MEXT) of Japan.

\clearpage


\begin{landscape}
\LongTables
\begin{deluxetable}{cccccccccccccccccccc}
\tabletypesize{\scriptsize}
\setlength{\tabcolsep}{0.01in}
\hspace{-3cm}
\tablecaption{Infrared and X-ray Properties of the AGNs in the 9-month {\it Swift}/BAT catalog
\label{tbl-1}}
\tablewidth{0pt}
\tablehead{
\colhead{No.} & \colhead{Object} &
 \colhead{$z$} &  \colhead{{\it f}(9)} & \colhead{{\it f}(18)} &  \colhead{{\it f}(90)} &
 \colhead{$\lambda L_{\lambda}$(9)} &
 \colhead{$\lambda L_{\lambda}$(18)} &
 \colhead{$\lambda L_{\lambda}$(90)} &
 \colhead{IR} &
 \colhead{$f_{\rm HX}$} &
 \colhead{$\log L_{\rm HX}$} & \colhead{$N_{\rm H}$} & 
 \colhead{$f_c$} & \colhead{X-ray}\\
 &&&&&&&&&\colhead{Catalog}&&&&&\colhead{Catalog}\\
\colhead{(1)} & \colhead{(2)} & \colhead{(3)} & \colhead{(4)} &
\colhead{(5)} & \colhead{(6)} & \colhead{(7)} &
\colhead{(8)} & \colhead{(9)} & \colhead{(10)} & 
\colhead{(11)} & \colhead{(12)} & \colhead{(13)} &
\colhead{(14)} & \colhead{(15)} \\
}
\startdata
1** & NGC 235A &  0.022229 &  0.136 &  0.295 &  \nodata &  43.60 &  43.74 &  \nodata &  (W, A, X) &  3.2 &  43.56 &  28.2 &  0.9975 & (a) \\
2** & Mrk 348 &  0.015034 &  0.308 &  0.593 &  0.736 &  43.54 &  43.69 &  43.09 &  (I, A, A) &  9.5 &  43.68 &  16.0 &  0.9960 & (c) \\
3 & Mrk 352 &  0.014864 &  0.012 &  0.016 &  \nodata &  42.20 &  42.02 &  \nodata &  (W, W, X) &  3.7 &  43.26 &  0.056 &  \nodata & (b) \\
4** & NGC 454 &  0.012125 &  0.213 &  0.417 &  \nodata &  43.19 &  43.16 &  \nodata &  (I, I, X) &  2.3 &  42.88 &  15.9 &  0.9970 & (a) \\
5 & Fairall 9 &  0.04702 &  0.229 &  0.440 &  \nodata &  44.59 &  44.58 &  \nodata &  (A, A, X) &  4.7 &  44.39 &  0.023 &  \nodata & (b) \\
6 & NGC 526A &  0.019097 &  0.141 &  0.292 &  \nodata &  43.58 &  43.60 &  \nodata &  (A, A, X) &  5.2 &  43.63 &  1.50 &  \nodata & (a) \\
7 & NGC 612 &  0.029771 &  0.135 &  0.153 &  2.604 &  43.96 &  43.52 &  44.25 &  (A, I, A) &  3.2 &  43.81 &  111 &  0.9945 & (a) \\
8** & ESO 297-G018 &  0.025201 &  0.081 &  0.132 &  0.672 &  43.49 &  43.41 &  43.51 &  (W, W, A) &  4.9 &  43.85 &  65.1 &  0.9972 & (a) \\
9 & NGC 788 &  0.013603 &  0.162 &  0.312 &  \nodata &  43.24 &  43.33 &  \nodata &  (W, A, X) &  5.9 &  43.39 &  46.9 &  0.9930 & (a) \\
10 & Mrk 1018 &  0.04244 &  0.059 &  0.085 &  \nodata &  43.81 &  43.69 &  \nodata &  (W, W, X) &  3.5 &  44.17 &  0 &  \nodata & (a) \\
11 & LEDA 138501 &  0.0492 &  \nodata &  \nodata &  \nodata &  \nodata &  \nodata &  \nodata &  (X, X, X) &  3.9 &  44.35 &  0.030 &  \nodata & (a) \\
12 & Mrk 590 &  0.02638 &  0.080 &  0.227 &  \nodata &  43.52 &  43.69 &  \nodata &  (W, W, X) &  3.7 &  43.77 &  0.027 &  \nodata & (b) \\
13 & 2MASX J02162987+5126246 &  0.0288 &  \nodata &  \nodata &  \nodata &  \nodata &  \nodata &  \nodata &  (X, X, X) &  3.6 &  43.84 &  1.74 &  \nodata & (a) \\
15 & NGC 931 &  0.016652 &  0.349 &  0.763 &  2.430 &  43.86 &  43.90 &  43.70 &  (A, A, A) &  7.3 &  43.66 &  0.360 &  \nodata & (a) \\
16 & NGC 985 &  0.043 &  0.165 &  0.368 &  1.291 &  44.37 &  44.42 &  44.27 &  (A, A, A) &  3.7 &  44.20 &  0.389 &  \nodata & (b) \\
17 & ESO 416-G002 &  0.059198 &  0.023 &  0.052 &  \nodata &  43.70 &  43.77 &  \nodata &  (W, W, X) &  3.2 &  44.43 &  0.027 &  \nodata & (b) \\
18 & ESO 198-024 &  0.0455 &  0.039 &  0.064 &  \nodata &  43.69 &  43.63 &  \nodata &  (W, W, X) &  3.9 &  44.28 &  0.100 &  \nodata & (b) \\
19 & QSO B0241+622 &  0.044 &  0.300 &  0.635 &  0.576 &  44.66 &  44.68 &  43.94 &  (A, A, A) &  7.3 &  44.52 &  0.742 &  \nodata & (a) \\
20** & NGC 1142 &  0.028847 &  0.265 &  0.380 &  \nodata &  44.22 &  44.08 &  \nodata &  (A, A, X) &  7.8 &  44.17 &  63.1 &  0.9974 & (e) \\
21 & 2MASX J03181899+6829322 &  0.0901 &  0.017 &  0.027 &  \nodata &  43.95 &  43.88 &  \nodata &  (W, W, X) &  3.5 &  44.85 &  4.10 &  0.9670 & (a) \\
22 & NGC 1275 &  0.017559 &  0.442 &  1.988 &  6.928 &  44.01 &  44.36 &  44.21 &  (A, A, A) &  11.5 &  43.90 &  0.151 &  \nodata & (a) \\
23 & PKS 0326--288 &  0.108 &  \nodata &  0.152 &  \nodata &  \nodata &  44.88 &  \nodata &  (X, A, X) &  2.3 &  44.84 &  0.031 &  \nodata & (a) \\
24 & NGC 1365 &  0.005457 &  2.234 &  5.364 &  80.384 &  43.70 &  43.78 &  44.25 &  (A, A, A) &  7.2 &  42.68 &  450 &  0.7400 & (f) \\
25 & ESO 548-G081 &  0.01448 &  0.248 &  0.097 &  0.968 &  43.42 &  42.68 &  43.18 &  (I, I, A) &  3.3 &  43.19 &  0 &  \nodata & (a) \\
27 & PGC 13946 &  0.036492 &  0.015 &  0.036 &  \nodata &  43.09 &  43.18 &  \nodata &  (W, W, X) &  2.9 &  43.95 &  85.0 &  0.9797 & (e) \\
28 & 2MASX J03565655--4041453 &  0.0747 &  0.020 &  0.048 &  \nodata &  43.85 &  43.95 &  \nodata &  (W, W, X) &  2.4 &  44.52 &  3.27 &  \nodata & (a) \\
29 & 3C 105 &  0.089 &  0.009 &  0.035 &  \nodata &  43.69 &  43.98 &  \nodata &  (W, W, X) &  3.4 &  44.83 &  28.2 &  0.9730 & (a) \\
30 & 3C 111.0 &  0.0485 &  0.081 &  0.135 &  \nodata &  44.07 &  44.01 &  \nodata &  (W, W, X) &  12.5 &  44.84 &  0.620 &  \nodata & (a) \\
31 & 1H 0419--577 &  0.104 &  0.081 &  0.106 &  \nodata &  44.70 &  44.68 &  \nodata &  (I, A, X) &  2.9 &  44.90 &  204 &  \nodata & (g) \\
32 & 3C 120 &  0.03301 &  0.203 &  0.497 &  1.468 &  44.23 &  44.31 &  44.08 &  (A, A, A) &  11.2 &  44.45 &  0.160 &  \nodata & (a) \\
33 & 2MASX J04440903+2813003 &  0.01127 &  0.090 &  0.145 &  1.427 &  42.93 &  42.75 &  43.13 &  (A, W, A) &  7.6 &  43.33 &  3.39 &  0.9900 & (a) \\
34 & MCG --01-13-025 &  0.015894 &  0.039 &  0.055 &  \nodata &  42.76 &  42.63 &  \nodata &  (W, W, X) &  4.5 &  43.41 &  0.004 &  \nodata & (b) \\
35 & 1RXS J045205.0+493248 &  0.029 &  \nodata &  \nodata &  \nodata &  \nodata &  \nodata &  \nodata &  (X, X, X) &  5.6 &  44.03 &  0.001 &  \nodata & (a) \\
36 & XSS J05054--2348 &  0.035043 &  0.060 &  0.124 &  \nodata &  43.65 &  43.68 &  \nodata &  (W, W, X) &  6.1 &  44.24 &  29.3 &  0.9914 & (e) \\
37 & 4U 0517+17 &  0.017879 &  0.166 &  0.748 &  0.957 &  43.60 &  43.95 &  43.36 &  (A, A, A) &  7.8 &  43.75 &  0 &  \nodata & (a) \\
38 & Ark 120 &  0.032296 &  0.252 &  0.253 &  \nodata &  44.30 &  44.00 &  \nodata &  (A, A, X) &  5.3 &  44.11 &  0.020 &  \nodata & (a) \\
39 & ESO 362-G018 &  0.012642 &  0.224 &  0.570 &  \nodata &  43.25 &  43.33 &  \nodata &  (I, I, X) &  5.1 &  43.26 &  26.6 &  0.9130 & (a) \\
40 & PICTOR A &  0.035058 &  0.073 &  0.135 &  \nodata &  43.73 &  43.72 &  \nodata &  (W, W, X) &  2.2 &  43.80 &  0.060 &  \nodata & (a) \\
45 & NGC 2110 &  0.007789 &  0.300 &  0.566 &  4.594 &  43.13 &  43.10 &  43.31 &  (A, A, A) &  25.6 &  43.54 &  2.84 &  0.9520 & (a) \\
46 & MCG +08-11-011 &  0.020484 &  0.340 &  1.283 &  2.377 &  44.03 &  44.30 &  43.87 &  (A, A, A) &  11.1 &  44.02 &  0.250 &  \nodata & (a) \\
47 & EXO 055620--3820.2 &  0.03387 &  0.532 &  0.691 &  \nodata &  44.50 &  44.29 &  \nodata &  (I, I, X) &  5.2 &  44.14 &  2.57 &  0.9660 & (a) \\
48 & IRAS 05589+2828 &  0.033 &  0.201 &  0.454 &  0.955 &  44.23 &  44.28 &  43.90 &  (A, A, A) &  5.6 &  44.15 &  0 &  \nodata & (a) \\
49** & ESO 005-G004 &  0.006228 &  0.537 &  0.520 &  8.501 &  43.18 &  42.86 &  43.38 &  (A, A, A) &  4.2 &  42.56 &  115 &  0.9973 & (e) \\
50 & Mrk 3 &  0.013509 &  0.322 &  1.892 &  2.939 &  43.64 &  44.10 &  43.60 &  (A, A, A) &  10.1 &  43.62 &  110 &  \nodata & (h) \\
51 & ESO 121-IG028 &  0.0403 &  0.016 &  0.037 &  \nodata &  43.20 &  43.27 &  \nodata &  (W, W, X) &  2.8 &  44.03 &  16.2 &  \nodata & (a) \\
52 & ESO 490-IG026 &  0.0248 &  0.173 &  0.706 &  \nodata &  43.73 &  44.02 &  \nodata &  (I, I, X) &  3.6 &  43.70 &  0.330 &  \nodata & (a) \\
53 & 2MASX J06403799--4321211 &  0.061 &  0.032 &  0.068 &  \nodata &  43.88 &  43.92 &  \nodata &  (W, W, X) &  2.8 &  44.40 &  16.1 &  \nodata & (a) \\
54 & 2MASX J06411806+3249313 &  0.047 &  0.042 &  0.089 &  \nodata &  43.75 &  43.80 &  \nodata &  (W, W, X) &  5.5 &  44.46 &  16.0 &  \nodata & (a) \\
55 & Mrk 6 &  0.01881 &  0.180 &  0.522 &  0.920 &  43.68 &  43.84 &  43.39 &  (A, A, A) &  6.6 &  43.72 &  3.26 &  0.9090 & (a) \\
56 & Mrk 79 &  0.022189 &  0.276 &  0.611 &  1.358 &  44.01 &  44.05 &  43.70 &  (A, A, A) &  4.7 &  43.72 &  0.006 &  \nodata & (b) \\
58 & IGR J07597--3842 &  0.04 &  \nodata &  \nodata &  \nodata &  \nodata &  \nodata &  \nodata &  (X, X, X) &  5.3 &  44.30 &  0 &  \nodata & (a) \\
60 & Mrk 18 &  0.011088 &  0.106 &  0.252 &  2.010 &  42.99 &  42.86 &  43.26 &  (A, I, A) &  3.1 &  42.93 &  18.2 &  0.9700 & (a) \\
61 & 2MASX J09043699+5536025 &  0.037 &  0.014 &  0.040 &  \nodata &  43.07 &  43.24 &  \nodata &  (W, W, X) &  3.4 &  44.03 &  0.060 &  \nodata & (a) \\
62 & 2MASX J09112999+4528060 &  0.026782 &  0.030 &  0.067 &  \nodata &  43.11 &  43.18 &  \nodata &  (W, W, X) &  3.0 &  43.69 &  33.0 &  0.9940 & (a) \\
63 & IRAS 09149--6206 &  0.0573 &  0.407 &  0.792 &  1.735 &  45.03 &  45.02 &  44.66 &  (A, A, A) &  3.2 &  44.40 &  0.850 &  \nodata & (a) \\
64 & 2MASX J09180027+0425066 &  0.156 &  0.021 &  0.050 &  \nodata &  44.55 &  44.66 &  \nodata &  (W, W, X) &  3.1 &  45.31 &  11.1 &  0.9870 & (a) \\
65 & MCG --01-24-012 &  0.019644 &  0.104 &  0.263 &  \nodata &  43.37 &  43.58 &  \nodata &  (W, A, X) &  4.6 &  43.60 &  11.4 &  \nodata & (a) \\
66 & MCG +04-22-042 &  0.032349 &  0.078 &  0.178 &  \nodata &  43.79 &  43.85 &  \nodata &  (A, A, X) &  4.1 &  44.00 &  0.039 &  \nodata & (b) \\
67 & Mrk 110 &  0.03529 &  0.073 &  0.107 &  \nodata &  43.74 &  43.62 &  \nodata &  (W, W, X) &  5.4 &  44.19 &  1.78 &  \nodata & (a) \\
68 & NGC 2992 &  0.007709 &  0.299 &  0.826 &  9.220 &  43.11 &  43.25 &  43.60 &  (A, A, A) &  6.6 &  42.94 &  1.19 &  0.4760 & (a) \\
69 & MCG --05-23-016 &  0.008486 &  0.384 &  1.391 &  1.277 &  43.31 &  43.57 &  42.83 &  (A, A, A) &  21.9 &  43.54 &  1.60 &  \nodata & (a) \\
70 & NGC 3081 &  0.007956 &  0.167 &  0.699 &  2.364 &  42.89 &  43.21 &  43.04 &  (A, A, A) &  8.8 &  43.09 &  98.0 &  0.9937 & (d) \\
71 & NGC 3227 &  0.003859 &  0.444 &  1.128 &  10.596 &  42.69 &  42.80 &  43.07 &  (A, A, A) &  12.9 &  42.63 &  1.74 &  0.8520 & (a) \\
72 & NGC 3281 &  0.010674 &  0.415 &  1.509 &  6.011 &  43.54 &  43.80 &  43.70 &  (A, A, A) &  7.3 &  43.27 &  86.3 &  0.9810 & (a) \\
73 & 2MASX J10384520--4946531 &  0.06 &  \nodata &  \nodata &  \nodata &  \nodata &  \nodata &  \nodata &  (X, X, X) &  3.3 &  44.45 &  1.81 &  0.8800 & (a) \\
74 & LEDA 093974 &  0.023923 &  0.096 &  0.256 &  0.941 &  43.62 &  43.74 &  43.61 &  (A, A, A) &  3.4 &  43.65 &  4.38 &  0.9860 & (a) \\
75** & Mrk 417 &  0.032756 &  0.068 &  0.152 &  \nodata &  43.64 &  43.71 &  \nodata &  (W, W, X) &  3.6 &  43.95 &  85.7 &  0.9983 & (a) \\
77 & NGC 3516 &  0.008836 &  0.262 &  0.651 &  1.317 &  43.17 &  43.27 &  42.87 &  (A, A, A) &  10.6 &  43.27 &  0.353 &  \nodata & (a) \\
78 & RX J1127.2+1909 &  0.1055 &  0.040 &  0.130 &  \nodata &  44.47 &  44.70 &  \nodata &  (W, W, X) &  2.2 &  44.80 &  0.270 &  \nodata & (a) \\
79 & NGC 3783 &  0.00973 &  0.502 &  1.530 &  2.716 &  43.54 &  43.72 &  43.27 &  (A, A, A) &  16.1 &  43.53 &  0.570 &  0.7220 & (a) \\
80 & SBS 1136+594 &  0.0601 &  0.041 &  0.083 &  \nodata &  43.97 &  43.99 &  \nodata &  (W, W, X) &  2.5 &  44.34 &  0.004 &  \nodata & (a) \\
81 & UGC 06728 &  0.006518 &  0.051 &  0.091 &  \nodata &  42.10 &  42.07 &  \nodata &  (W, W, X) &  3.7 &  42.54 &  0.010 &  \nodata & (a) \\
82 & 2MASX J11454045--1827149 &  0.032949 &  0.078 &  0.131 &  \nodata &  43.70 &  43.65 &  \nodata &  (W, W, X) &  3.9 &  43.99 &  0.035 &  \nodata & (a) \\
83 & CGCG 041--020 &  0.036045 &  0.054 &  0.112 &  \nodata &  43.62 &  43.66 &  \nodata &  (W, W, X) &  2.5 &  43.88 &  10.8 &  0.9910 & (a) \\
84 & IGR J12026--5349 &  0.027966 &  0.166 &  0.614 &  1.630 &  44.00 &  44.26 &  43.99 &  (A, A, A) &  4.0 &  43.86 &  2.34 &  \nodata & (a) \\
85 & NGC 4051 &  0.002335 &  0.346 &  0.885 &  4.557 &  42.12 &  42.23 &  42.24 &  (A, A, A) &  4.6 &  41.74 &  0.029 &  \nodata & (b) \\
86** & Ark 347 &  0.02244 &  0.091 &  0.102 &  \nodata &  43.43 &  43.29 &  \nodata &  (W, A, X) &  2.3 &  43.42 &  30.0 &  0.9950 & (a) \\
87 & NGC 4102 &  0.002823 &  1.082 &  3.287 &  54.050 &  42.80 &  42.98 &  43.50 &  (A, A, A) &  2.4 &  41.62 &  210 &  \nodata & (i) \\
88 & NGC 4138 &  0.002962 &  0.044 &  0.075 &  2.161 &  41.34 &  41.29 &  42.15 &  (W, W, A) &  2.1 &  41.61 &  8.00 &  0.9880 & (a) \\
89 & NGC 4151 &  0.003319 &  1.032 &  3.629 &  4.594 &  42.91 &  43.16 &  42.56 &  (A, A, A) &  37.4 &  42.96 &  5.32 &  0.9590 & (a) \\
90 & Mrk 766 &  0.012929 &  0.220 &  0.859 &  3.312 &  43.43 &  43.72 &  43.61 &  (A, A, A) &  2.3 &  42.93 &  0.525 &  \nodata & (b) \\
91 & NGC 4388 &  0.008419 &  0.462 &  1.589 &  10.349 &  43.38 &  43.61 &  43.73 &  (A, A, A) &  25.3 &  43.60 &  36.2 &  0.9920 & (a) \\
92 & NGC 4395 &  0.001064 &  0.013 &  0.052 &  \nodata &  39.94 &  40.25 &  \nodata &  (W, W, X) &  2.6 &  40.81 &  3.30 &  0.6780 & (a) \\
94 & NGC 4507 &  0.011802 &  0.510 &  1.163 &  4.370 &  43.72 &  43.78 &  43.65 &  (A, A, A) &  19.3 &  43.78 &  34.3 &  0.9710 & (a) \\
95** & ESO 506-G027 &  0.025024 &  0.114 &  0.207 &  0.613 &  43.73 &  43.69 &  43.46 &  (A, A, A) &  13.2 &  44.28 &  84.1 &  0.9981 & (j) \\
96 & XSS J12389--1614 &  0.036675 &  0.053 &  0.109 &  \nodata &  43.63 &  43.67 &  \nodata &  (W, W, X) &  5.8 &  44.26 &  3.25 &  \nodata & (a) \\
97 & NGC 4593 &  0.009 &  0.344 &  0.569 &  \nodata &  43.14 &  43.23 &  \nodata &  (I, A, X) &  9.1 &  43.21 &  0.031 &  \nodata & (a) \\
98 & WKK 1263 &  0.02443 &  0.091 &  0.170 &  0.769 &  43.51 &  43.58 &  43.54 &  (W, A, A) &  2.8 &  43.58 &  0.060 &  \nodata & (a) \\
100 & SBS 1301+540 &  0.02988 &  0.015 &  0.022 &  \nodata &  42.91 &  42.78 &  \nodata &  (W, W, X) &  2.5 &  43.71 &  0.040 &  \nodata & (b) \\
101 & NGC 4945 &  0.001878 &  8.811 &  9.945 &  \nodata &  43.36 &  43.11 &  \nodata &  (A, A, X) &  19.4 &  42.18 &  500 &  \nodata & (k) \\
102** & NGC 4992 &  0.025137 &  0.059 &  \nodata &  \nodata &  43.45 &  \nodata &  \nodata &  (A, X, X) &  4.7 &  43.83 &  69.0 &  0.9974 & (a) \\
103 & MCG --03-34-064 &  0.016541 &  0.453 &  1.873 &  4.634 &  43.96 &  44.28 &  43.97 &  (A, A, A) &  4.7 &  43.46 &  40.7 &  0.9610 & (a) \\
104 & Cen A &  0.001825 &  10.190 &  13.150 &  \nodata &  43.40 &  43.20 &  \nodata &  (A, A, X) &  74.8 &  42.74 &  5.50 &  \nodata & (a) \\
105 & MCG --06-30-015 &  0.007749 &  0.280 &  0.591 &  1.035 &  43.08 &  43.11 &  42.65 &  (A, A, A) &  7.5 &  43.00 &  0.190 &  \nodata & (a) \\
106 & NGC 5252 &  0.022975 &  0.104 &  0.136 &  0.416 &  43.51 &  43.35 &  43.22 &  (W, W, A) &  6.6 &  43.90 &  4.34 &  0.9620 & (a) \\
107 & 4U 1344-60 &  0.012879 &  0.207 &  0.556 &  \nodata &  43.41 &  43.54 &  \nodata &  (A, A, X) &  7.0 &  43.41 &  2.50 &  \nodata & (a) \\
108 & IC 4329A &  0.016054 &  0.769 &  1.790 &  1.785 &  44.16 &  44.23 &  43.53 &  (A, A, A) &  30.0 &  44.24 &  0.610 &  \nodata & (a) \\
109 & Mrk 279 &  0.030451 &  0.141 &  0.387 &  \nodata &  43.99 &  44.13 &  \nodata &  (A, A, X) &  4.4 &  43.97 &  0.013 &  \nodata & (a) \\
110 & NGC 5506 &  0.006181 &  0.823 &  2.240 &  8.413 &  43.36 &  43.50 &  43.37 &  (A, A, A) &  23.6 &  43.30 &  2.78 &  0.9893 & (a) \\
112 & NGC 5548 &  0.01717 &  0.157 &  0.409 &  1.073 &  43.54 &  43.65 &  43.37 &  (A, A, A) &  5.8 &  43.59 &  0.070 &  \nodata & (a) \\
113 & ESO 511-G030 &  0.02239 &  0.064 &  0.090 &  0.847 &  43.27 &  43.14 &  43.50 &  (W, W, A) &  4.7 &  43.73 &  0.098 &  \nodata & (a) \\
115 & NGC 5728 &  0.0093 &  0.176 &  0.418 &  11.383 &  43.05 &  43.12 &  43.86 &  (A, A, A) &  8.9 &  43.23 &  82.0 &  \nodata & (a) \\
116 & Mrk 841 &  0.036422 &  0.126 &  0.372 &  \nodata &  44.11 &  44.27 &  \nodata &  (A, A, X) &  5.1 &  44.20 &  0.219 &  \nodata & (b) \\
117 & Mrk 290 &  0.029577 &  0.085 &  0.151 &  \nodata &  43.65 &  43.70 &  \nodata &  (W, A, X) &  3.0 &  43.78 &  0.150 &  \nodata & (a) \\
118 & Mrk 1498 &  0.0547 &  0.067 &  0.214 &  \nodata &  44.20 &  44.40 &  \nodata &  (A, A, X) &  4.5 &  44.51 &  58.0 &  0.9905 & (a) \\
119 & 2MASX J16481523--3035037 &  0.031 &  0.030 &  0.038 &  \nodata &  43.23 &  43.06 &  \nodata &  (W, W, X) &  8.6 &  44.28 &  0.240 &  \nodata & (a) \\
120 & NGC 6240 &  0.02448 &  0.350 &  1.489 &  23.100 &  44.20 &  44.53 &  45.02 &  (A, A, A) &  4.7 &  43.81 &  102 &  \nodata & (a) \\
122 & NGC 6300 &  0.003699 &  0.277 &  1.336 &  14.928 &  42.44 &  42.82 &  43.17 &  (A, A, A) &  9.1 &  42.44 &  21.5 &  \nodata & (a) \\
123 & GRS 1734--292 &  0.0214 &  \nodata &  \nodata &  \nodata &  \nodata &  \nodata &  \nodata &  (X, X, X) &  10.9 &  44.05 &  0.912 &  \nodata & (a) \\
124 & 1RXS J174538.1+290823 &  0.111332 &  0.030 &  0.059 &  \nodata &  44.40 &  44.41 &  \nodata &  (W, W, X) &  3.9 &  45.09 &  0.139 &  \nodata & (a) \\
125 & 3C 382 &  0.05787 &  0.120 &  0.106 &  \nodata &  44.50 &  43.95 &  \nodata &  (A, I, X) &  8.1 &  44.81 &  0.013 &  \nodata & (b) \\
126** & ESO 103-035 &  0.013286 &  0.300 &  1.446 &  1.227 &  43.59 &  43.97 &  43.20 &  (A, A, A) &  9.7 &  43.58 &  21.6 &  0.9990 & (a) \\
127 & 3C 390.3 &  0.0561 &  0.090 &  0.242 &  \nodata &  44.35 &  44.48 &  \nodata &  (A, A, X) &  10.1 &  44.88 &  0.120 &  \nodata & (a) \\
128 & NVSS J193013+341047 &  0.0629 &  0.130 &  0.254 &  \nodata &  44.62 &  44.61 &  \nodata &  (A, A, X) &  3.3 &  44.50 &  27.6 &  0.9260 & (a) \\
129 & NGC 6814 &  0.005214 &  0.334 &  0.258 &  6.954 &  42.65 &  42.41 &  43.14 &  (I, A, A) &  6.2 &  42.57 &  0.058 &  \nodata & (b) \\
130 & 3C 403 &  0.059 &  0.093 &  0.215 &  \nodata &  44.31 &  44.39 &  \nodata &  (W, W, X) &  4.1 &  44.53 &  45.0 &  \nodata & (l) \\
131 & Cyg A &  0.05607 &  0.152 &  0.418 &  2.455 &  44.47 &  44.72 &  44.79 &  (W, A, A) &  10.9 &  44.91 &  11.0 &  \nodata & (a) \\
133 & NGC 6860 &  0.014884 &  0.155 &  0.357 &  1.369 &  43.41 &  43.47 &  43.35 &  (A, A, A) &  4.9 &  43.39 &  0.100 &  \nodata & (m) \\
136 & 4C +74.26 &  0.104 &  0.147 &  0.175 &  \nodata &  45.13 &  44.90 &  \nodata &  (A, A, X) &  5.0 &  45.14 &  0.177 &  \nodata & (n) \\
137 & Mrk 509 &  0.0344 &  0.247 &  0.499 &  \nodata &  44.35 &  44.35 &  \nodata &  (A, A, X) &  9.7 &  44.42 &  0.015 &  \nodata & (a) \\
138 & IC 5063 &  0.011348 &  1.159 &  2.246 &  3.821 &  43.87 &  44.03 &  43.56 &  (I, A, A) &  7.1 &  43.31 &  25.0 &  0.9910 & (g) \\
139 & 2MASX J21140128+8204483 &  0.084 &  0.071 &  0.105 &  \nodata &  44.62 &  44.48 &  \nodata &  (A, A, X) &  3.6 &  44.80 &  0.121 &  \nodata & (a) \\
140 & IGR J21247+5058 &  0.02 &  \nodata &  \nodata &  \nodata &  \nodata &  \nodata &  \nodata &  (X, X, X) &  13.9 &  44.10 &  2.47 &  \nodata & (a) \\
141 & IGR J21277+5656 &  0.0147 &  0.211 &  0.435 &  \nodata &  43.53 &  43.55 &  \nodata &  (A, A, X) &  2.7 &  43.12 &  1.23 &  \nodata & (a) \\
142 & RX J2135.9+4728 &  0.025 &  \nodata &  0.240 &  \nodata &  \nodata &  43.76 &  \nodata &  (X, A, X) &  2.9 &  43.62 &  0.825 &  \nodata & (a) \\
144 & UGC 11871 &  0.026612 &  0.146 &  0.320 &  5.040 &  43.90 &  43.94 &  44.43 &  (A, A, A) &  3.9 &  43.80 &  2.44 &  \nodata & (a) \\
145** & NGC 7172 &  0.008683 &  0.316 &  0.424 &  8.087 &  43.24 &  43.07 &  43.65 &  (A, A, A) &  12.4 &  43.32 &  8.19 &  0.9990 & (a) \\
146 & NGC 7213 &  0.005839 &  0.360 &  0.743 &  2.943 &  42.95 &  42.77 &  42.86 &  (A, I, A) &  5.2 &  42.59 &  0.025 &  \nodata & (a) \\
147 & NGC 7314 &  0.00476 &  0.268 &  0.304 &  4.499 &  42.48 &  42.41 &  42.88 &  (I, A, A) &  5.7 &  42.46 &  1.16 &  \nodata & (a) \\
148** & NGC 7319 &  0.022507 &  0.088 &  0.158 &  0.576 &  43.53 &  43.48 &  43.34 &  (A, A, A) &  4.1 &  43.67 &  86.6 &  0.9965 & (a) \\
149 & 3C 452 &  0.0811 &  0.029 &  0.070 &  \nodata &  44.09 &  44.19 &  \nodata &  (W, W, X) &  3.3 &  44.73 &  23.0 &  0.9360 & (a) \\
151 & MR 2251--178 &  0.06398 &  0.090 &  0.148 &  \nodata &  44.36 &  44.30 &  \nodata &  (W, W, X) &  10.8 &  45.03 &  0.280 &  \nodata & (a) \\
152 & NGC 7469 &  0.016317 &  0.767 &  2.692 &  27.694 &  44.18 &  44.42 &  44.74 &  (A, A, A) &  8.3 &  43.70 &  0.041 &  \nodata & (b) \\
153 & Mrk 926 &  0.04686 &  0.060 &  0.214 &  0.647 &  44.01 &  44.26 &  44.04 &  (A, A, A) &  5.5 &  44.45 &  0.035 &  \nodata & (a) \\
154 & NGC 7582 &  0.005254 &  1.368 &  3.287 &  60.906 &  43.43 &  43.51 &  44.08 &  (A, A, A) &  6.7 &  42.61 &  33.0 &  \nodata & (o)
\enddata

\tablecomments{
Tables~\ref{tbl-1} summarize the infrared to X-ray properties of all the
135 Swift/BAT 9 month non-blazar AGNs in \cite{tue08}, including 128
 with MIR counterparts: (1) source No.\ in \citep{tue08}.
The sources with asterisks** represent ``new type'' AGNs which exhibit extremely small scattered fraction
($f_{\rm scat} \equiv 1-f_{\rm c} \leq 0.005$, thus $f_{\rm c} \geq 0.995$) suggesting the geometrically thick
tori around the nuclei.  (2) object name, (3)
redshift, (4)--(6) infrared fluxes ($F_\nu$) at 9 $\mu$m, 18
$\mu$m, and 90 $\mu$m in units of Jansky (Jy), (7)--(9) infrared luminosities ($\lambda
L_\lambda$) at 9 $\mu$m, 18 $\mu$m, and 90 $\mu$m in units of erg s$^{-1}$, (10) IR reference
catalogs for 9 $\mu$m, 18 $\mu$m, and 90 $\mu$m, (11)
hard X-ray flux in the 14--195 keV band in units of $10^{-11}$erg s$^{-1}$cm$^{-2}$, (12) hard X-ray luminosity in
the 14--195 keV band $(\log L_{\rm HX})$ in units of erg s$^{-1}$, (13) X-ray absorption column density ($N_{\rm 
H}$), (14) covering fraction in the X-ray spectrum ($f_{\rm c}$), and
(15) the reference for the X-ray spectra (a): \cite{win09a} (b): \cite{tue08} (c): \cite{nog10} (d): \cite{egu11}
(e): \cite{egu09} (f): \cite{ris09} (g): \cite{tur09} (h): \cite{awa08} (i): \cite{gon11} (j): \cite{win09b} (k): \cite{ito08} 
(l)\cite{taz11} (m):  \cite{win10} (n): \cite{bal05} (o): \cite{bia09}.
 For AGNs whose AKARI MIR
fluxes are not available, we convert the infrared fluxes and luminosities with IRAS or
WISE into those at 9 $\mu$m or 18 $\mu$m according to the formula given
in Section~2.5. Columns (1), (2), (3), (11) are taken from
\cite{tue08} except for the redshifts of two sources,
 No.\ 13 (2MASX J02162987+5126246) and No.\ 53 (2MASX
 J06403799--4321211), which are adopted from the Swift/BAT 58 month catalog
(\url{http://heasarc.gsfc.nasa.gov/docs/swift/results/bs58mon/}). 
The X-ray spectral information (columns 13--14) is basically adopted from \cite{win09a}, while we refer to the results
obtained with {\it Suzaku} \citep{egu09, egu11, ris09, tur09, awa08, gon11, win09b, ito08, taz11, win10, bia09} and those with {\it
XMM-Newton} \citep{nog10, bal05} whenever available. When the information is not available in \cite{win09a}, we refer to \cite{tue08}.
  All luminosities in this table are calculated by using the redshift given in column (3).
 }\\
\end{deluxetable}

\clearpage
\end{landscape}

\begin{deluxetable}{cc*{6}c}
\tabletypesize{\scriptsize}
\tablecaption{Correlation parameters between MIR and the hard X-ray sample
\label{tbl-2}}
\tablewidth{0pt}
\tablehead{
\colhead{Sample} & \colhead{$N$} & \colhead{$\rho_{L}$} & \colhead{$\rho_{f}$} & \colhead{$P_L$} & \colhead{$P_f$} & \colhead{$a$} &
 \colhead{$b$}
 \\
\colhead{(1)} & \colhead{(2)} & \colhead{(3)} & \colhead{(4)} &
\colhead{(5)} & \colhead{(6)} & \colhead{(7)} & \colhead{(8)} 
}
\startdata
9 $\mu$m		& 126 & 0.82 & 0.60 & $3.0\times10^{-31}$ & $1.4\times10^{-13}$ &  $0.06\pm0.07$ &$ 1.12\pm0.08$ \\
18 $\mu$m 	& 127 & 0.76 & 0.59 & $1.7\times10^{-25}$ & $2.4\times10^{-13}$ &$ 0.02\pm0.07 $&$ 1.10\pm0.07 $   \\
90 $\mu$m	&     62 & 0.59 & 0.17 & $4.5\times10^{-7}$ & $1.8\times10^{-1}$ & $-0.21\pm0.10 $& $1.16\pm0.11$
\enddata
\tablecomments{Correlation Proporties between 14-195 keV X-ray luminosity ($\log L_{\rm HX}$)
 and  infrared (9, 18, and 90 $\mu$m) luminosities ($\log \lambda L_{\lambda}(9, 18, 90~\mu{\rm m})$) to various subsample populations.
 (1) sample name; (2) number of sample; (3) the Spearman's Rank coefficient for luminosity correlations ($\rho_L$)
; (4) the Spearman's Rank coefficient for flux-flux correlations ($\rho_f$); (5) the standard Student $t$-test null significance
 level for luminosity correlations ($P_L$); (6) the standard Student $t$-test null significance level for flux-flux correlations
  ($P_f$); (7) regression intercept ($a$) and its 1$\sigma$ uncertainty; (8) slope value ($b$) and its 1$\sigma$ uncertainty.
Equation is represented as $Y=a+bX$.
}\\
\end{deluxetable}

\begin{deluxetable}{lccc}
\tabletypesize{\scriptsize}
\tablecaption{average and standard deviation of $\log  (\lambda L_{\lambda}(9, 18 \mu {\rm m})/{L_{\rm HX}})$
\label{tbl-3}}
\tablewidth{0pt}
\tablehead{
\colhead{Sample} & \colhead{$N$} & \colhead{$\bar{r}$} & \colhead{$\sigma$} \\
\colhead{(1)} & \colhead{(2)} & \colhead{(3)} & \colhead{(4)}
}
\startdata
9 $\mu$m All & 126 & $-0.129 \pm 0.039$ & $0.437 \pm 0.055$  \\
Type-1 & 55 & $-0.165 \pm 0.052$ & $0.389 \pm 0.074$  \\
Type-2($+$ New Type) & 71 & $-0.101 \pm 0.056$ & $0.470 \pm 0.079$  \\
18 $\mu$m All & 127 & $-0.080 \pm 0.042$ & $0.473 \pm 0.059$  \\
Type-1 & 57 & $-0.141 \pm 0.060$ & $0.452 \pm 0.085$  \\
Type-2($+$ New Type) & 70 & $-0.031 \pm 0.058$ & $0.483 \pm 0.082$  
\enddata
\tablecomments{Averages and standard deviations of $\log (\lambda L_{\lambda}(9, 18~\mu{\rm m})/{L_{\rm HX}})$.
(1) Sample name; (2) Number of sample; (3) average of $\log  (\lambda L_{\lambda}(9, 18~\mu {\rm m})/{L_{\rm HX}})$; (4) standard deviation of $\log  (\lambda L_{\lambda}(9, 18~\mu {\rm m})/{L_{\rm HX}})$}\\
\end{deluxetable}

\end{document}